
\input harvmac.tex
\input epsf.tex


%
%
%

%
%
\def\dag{\dagger}

\def\tilde{\widetilde}
\def\bar{\overline}
\def\hat{\widehat}
\def\*{\star}
\def\[{\left[}
\def\]{\right]}
\def\({\left(}		
\def\){\right)}

%
%
\def\zb{{\bar{z} }}
\def\frac#1#2{{#1 \over #2}}
\def\inv#1{{1 \over #1}}
\def\half{{1 \over 2}}

\def\2pi{\hbox{$2\pi i$}}

\def\dsl{\raise.15ex\hbox{/}\kern-.57em\partial}
\def\Dsl{\,\raise.15ex\hbox{/}\mkern-.13.5mu D}
%
%
\def\th{\theta}		\def\Th{\Theta}

\def\al{\alpha}
\def\ep{\epsilon}

%
%

%

\def\2pi{\hbox{$2\pi i$}}

\def\dsl{\raise.15ex\hbox{/}\kern-.57em\partial}
\def\Dsl{\,\raise.15ex\hbox{/}\mkern-.13.5mu D}
%
%
%
\font\numbers=cmss12
\font\upright=cmu10 scaled\magstep1
\def\stroke{\vrule height8pt width0.4pt depth-0.1pt}
\def\topfleck{\vrule height8pt width0.5pt depth-5.9pt}
\def\totfleck{\vrule height2pt width0.5pt depth0.1pt}
\def\Zmath{\vcenter{\hbox{\numbers\rlap{\rlap{Z}\kern 0.8pt\topfleck}\kern
2.2pt
                   \rlap Z\kern 6pt\totfleck\kern 1pt}}}
\def\Qmath{\vcenter{\hbox{\upright\rlap{\rlap{Q}\kern
                   3.8pt\stroke}\phantom{Q}}}}
\def\Nmath{\vcenter{\hbox{\upright\rlap{I}\kern 1.7pt N}}}
\def\Cmath{\vcenter{\hbox{\upright\rlap{\rlap{C}\kern
                   3.8pt\stroke}\phantom{C}}}}
\def\Rmath{\vcenter{\hbox{\upright\rlap{I}\kern 1.7pt R}}}
\def\Z{\ifmmode\Zmath\else$\Zmath$\fi}
\def\Q{\ifmmode\Qmath\else$\Qmath$\fi}
\def\N{\ifmmode\Nmath\else$\Nmath$\fi}
\def\C{\ifmmode\Cmath\else$\Cmath$\fi}
\def\R{\ifmmode\Rmath\else$\Rmath$\fi}

\lref\Affrev{For a review, see I. Affleck, ``Conformal Field Theory
Approach to
Quantum Impurity Problems'', UBCTP-93-25, cond-mat/9311054.}
\lref\AFL{N. Andrei, K. Furuya, and J. Lowenstein, Rev. Mod. Phys. 55
(1983) 331; \hfill\break
A.M. Tsvelick and P.B. Wiegmann, Adv. Phys. 32 (1983) 453.}
\lref\RS{N. Reshetikhin and H. Saleur, Nucl. Phys. B419 (1994) 507.}
\lref\KR{A.N. Kirillov and N. Reshetikhin, J. Phys. A20 (1987) 1565,
1587.}
\lref\TS{M. Takahashi and M. Suzuki, Prog. Th. Phys. 48 (1972) 2187.}
\lref\chered{I. Cherednik, Theor. Math. Phys. 61 (1984) 977.}
\lref\GZ{S. Ghoshal and A.B. Zamolodchikov,
 Int. J. Mod. Phys. A9 (1994) 3841.}
\lref\fki{A. Fring and R. K\"oberle, ``Factorized Scattering in the
Presence of
Reflecting Boundaries'', USP-IFQSC/TH/93-06, hep-th/9304141.}
\lref\fkii{A. Fring and R. K\"oberle, ``Affine Toda Field Theory in
the
Presence of  Reflecting Boundaries'', USP-IFQSC/TH/93-12,
hep-th/9309142.}
\lref\ghosi{S. Ghoshal, ``Bound State Boundary $S$ Matrix of the
Sine-Gordon
Model'',  RU-93-51, hep-th/9310188.}
\lref\ghosii{S. Ghoshal, ``Boundary $S$ Matrix of the $O(n)$
Symmetric
Nonlinear Sigma Model'' RU-94-02, hep-th/9401008.}
\lref\sasaki{R. Sasaki, ``Reflection Bootstrap Equations for Toda
Field
Theory'', 
 hep-th/9311027.}
\lref\ABBBQ{F. Alcaraz, M. Barber, M. Batchelor, R. Baxter and G.
Quispel, J.
Phys. A20 (1987) 6397.}
\lref\ressmir{N. Reshetikhin and F. Smirnov, Commun. Math. Phys. 131
(1990), 157}
\lref\skly{E.K. Sklyanin, J. Phys. A21 (1988) 2375.}
\lref\ZandZ{A.B. Zamolodchikov and Al.B. Zamolodchikov, Ann.  Phys.
120 (1980) 253.}
\lref\AL{I. Affleck and A. Ludwig, Phys. Rev. Lett. 67 (1991) 161.}
\lref\Cardybc{J.L. Cardy, Nucl. Phys. B275 [FS17] (1986), 200.}
\lref\CarVer{J. Cardy, Nucl. Phys. B324 (1989) 581.}
\lref\EA{S. Eggert and I. Affleck, Phys. Rev. B46 (1992) 10866.}
\lref\pkondo{P. Fendley, Phys. Rev. Lett. 71 (1993) 2485,
cond-mat/9304031.}
\lref\FS{P. Fendley and H. Saleur, Nucl. Phys. B388 (1992) 609,
hep-th/9204094.}
\lref\DL{C. Destri and J.H. Lowenstein, Nucl. Phys. B205 (1982) 369.}
\lref\DNW{G.I. Dzapardize,  A.A. Nersesyan and P.B. Wiegmann, Phys.
Scr. 27
(1983) 5.}
\lref\DV{C. Destri and H. de Vega, J. Phys. A22 (1989) 1329.}
\lref\toundrefs{L. Mezincescu and R.I. Nepomechie, Int. J. Mod. Phys.
A6 (1991)
 5231; Int. J. Mod. Phys. A7 (1992) 565; H.J. de Vega and A.
Gonzalez-Ruiz,
 preprint LPTHE-93/38.}
\lref\EKS{F. Essler, V. Korepin and K. Schoutens, J. Phys A25 (1992)
4115.}
\lref\boundrefs{L. Mezincescu and R.I. Nepomechie, Int. J. Mod. Phys. A6 (1991)
 5231; Int. J. Mod. Phys. A7 (1992) 565; H.J. de Vega and A. Gonzalez-Ruiz,
 preprint LPTHE-93/38.}
\lref\nonstrings{F. Woynarovich, J. Phys. A15 (1982) 2985;\hfill\break
O. Babelon, H.J. de Vega and C.M. Viallet, Nucl. Phys. B220
 (1983) 13.}
\lref\FSW{P. Fendley, H. Saleur, N. P. Warner,
Nucl. Phys. B430 (1994) 577.}
\lref\DDV{C.Destri, H.de Vega, Phys. Rev. Lett. 69 (1992) 2313.}
\lref\FSZ{P. Fendley, H. Saleur, Al. Zamolodchikov, Int. J. Mod. Phys. A32
(1993) 5752; 5717.}
\lref\rcardy{J. Cardy,
Nucl. Phys. B240 (1984) 514; Nucl. Phys. B324 (1989) 581.}
\lref\rtba{Al. B. Zamolodchikov, Nucl. Phys. B342 (1990) 695.}
\lref\rchat{R. Chatterjee, {\it Exact Partition
Function and Boundary State of Critical Ising Model with Boundary
Magnetic Field}, Rutgers preprint RU-94-95}
\lref\rbaz{V. V. Bazhanov, S. L. Lukyanov, and A. B.
Zamolodchikov, {\it Integrable Structure of Conformal Field Theory,
Quantum KdV Theory and Thermodynamic Bethe Ansatz}, Cornell
and Rutgers preprints CLNS 94/1316, RU-94-98, hep-th/9412229}
\lref\gins{
P. Ginsparg, Les Houches 1988 Lectures, E. Br\'ezin and J. Zinn-Justin,
Eds.}
\lref\FS{P. Fendley, H. Saleur, Nucl. Phys.
B428 (1994) 681.}
\lref\ALZunpub{Al. Zamolodchikov, {\it Mass scale in sine-Gordon and its
reductions}, unpublished manuscript.}
\lref\DV{C. Destri, H. de Vega, Nucl. Phys. B358 (1991) 251.}
\lref\ZZ{A. B. Zamolodchikov,
Al. B. Zamolodchikov, Annals of Physics, 120 (1979) 253.}
\lref\AlZ{Al. Zamolodchikov, Nucl. Phys. B432 (1994) 427.}
\lref\KR{A. N. Kirillov, N. Yu Reshetikhin, J. Phys. A20 (1987) 1565; 1587.}
\lref\BS{M. Bauer, H. Saleur, Nucl. Phys. B320 (1989) 591.}
\lref\PS{V. Pasquier, H. Saleur, Nucl. Phys. B330 (1990) 523.}
\lref\Sasha{A. B.
Zamolodchikov, in Proceedings of {\it Quantum field theory, statistical
mechanics, quantum groups and topology}, Miami 1991, Eds. T. Curtright et al.,
Worlds Scientific.}
\lref\SS{S. Skorik, H. Saleur, {\it Boundary bound states and boundary
bootstrap in the sine-gordon model with Dirichlet boundary conditions}, USC
preprint USC-95-01, hep-th/9502011.}
\lref\NM{L. Mezincescu, R. Nepomechie, J.Phys.A25 (1992) 2533.}
\lref\GNM{M. T. Grisaru, L. Mezincescu, R. Nepomechie, {\it Direct calculation
of the boundary S matrix for the open Heisenberg chain}, preprint UMTG-177,
hep-th/9407089.}

\def\t{\theta}

\def\<{\langle}
\def\>{\rangle}
\noblackbox
\Title{\vbox{\baselineskip12pt
\hbox{CLNS/95-1328,  USC-95-005}
\hbox{ISAS/EP/95-26}
\hbox{hep-th/9503227}}}
{\vbox{\centerline{Boundary energy and boundary states}
\vskip4pt\centerline{ in integrable quantum field theories
}}}

\centerline{A. LeClair $^\heartsuit$,
G. Mussardo $^\diamondsuit$, H.
Saleur$^\spadesuit$ and S. Skorik $^\dagger$ }
\vskip2pt
\centerline{$^\heartsuit$ Newman Laboratory, Cornell University}
\centerline{Ithaca, NY 14853}
\centerline{$^\diamondsuit$ International School for Advanced Studies
and INFN}
\centerline{34014 Trieste}
\vskip2pt
\centerline{$^\dagger$Department of Physics, University of Southern
California}
\centerline{Los Angeles CA 90089}
\vskip2pt
\centerline{$^\spadesuit$Department of Physics and Department of
Mathematics}
\centerline{University of Southern California}
\centerline{Los Angeles CA 90089}
\vskip.3in

{\bf Abstract}: We study the ground state energy of integrable $1+1$ quantum
field theories with boundaries (the genuine Casimir effect). In the scalar
case, this is done by introducing a new, ``R-channel TBA'', where the boundary
is represented by a boundary state,
and the thermodynamics involves evaluating scalar products of  boundary states
with all the states of the theory. In the non-scalar, sine-Gordon case, this is
done by generalizing the method of Destri and De Vega. The two approaches are
compared. Miscellaneous other results are obtained, in particular formulas for
the
overall normalization and scalar products of boundary states,  exact
partition functions for the critical Ising model in a boundary magnetic
field, and also results
for the energy, excited states and boundary S-matrix of $O(n)$ and minimal
models.

\Date{3/95}

\newsec{Introduction}

The main  purpose of this paper is to study the ground state energy of 1+1
integrable relativistic quantum field theories with boundaries. This involves
several questions. One is the energy associated with a boundary for an
infinite system, the other is the way the energy of the theory on an interval
varies with its length - the ``genuine'' Casimir effect.

Some of these properties are most easily studied directly in a continuum
formalism, using a scattering description for either a massive or massless
bulk theory, the information on the boundary being encoded into a reflection
matrix. Other properties
are most easily studied using a lattice regularization.

In computing energies, we shall use detailed information about the boundary
states, and this work also presents new results on that question. For
instance, the ``R channel'' of the Thermodynamical Bethe Ansatz (TBA) approach
provides expressions for the scalar products of boundary states.

For a Quantum Field Theory defined on a torus, the standard way to
compute its ground state energy is through the Thermodynamic Bethe
Ansatz \rtba. If the theory is defined on a circle of circumference $R$,
one switches to a modular transformed point of view where now the theory is
defined on a circle of very long circumference $L$ and at temperature $1/R$.
The free energy $F$ of the theory in the ``$R$ channel'' can be computed
using TBA, and it is simply related to the ground state energy $E^0(R)$ of
the theory in the ``$L$ channel'' by $F=-TLE^0(R)$.

Consider now a quantum field theory on a cylinder of finite length
$R$ and circumference $L$, with some boundary conditions $(a,b)$
at the ends of the cylinder.  There are two possible ways of viewing
the partition function:  in the `R-channel' where time evolution is
in the R direction, and in the `L-channel' where time evolution is
in the L direction.  (See figure 1).
This problem is more difficult because the
theory looks rather different in the two channels. To use the same argument
and do a TBA in the R channel, one now needs to control the effect of boundary
states on the thermodynamics. This is discussed in section 2 for the simple
case of the Ising model to clarify a few delicate points, and in section 3
for the general case of a scalar theory. The TBA in the L channel gives rise
to a different kind of information, related to the normalization of boundary
states. This is discussed in section 4. Finally, all these results can be
generalized to the case where the theory is still conformal in the bulk but
has a boundary interaction that breaks the conformal invariance (section
5).

\midinsert
\epsfxsize = 3in
\bigskip\bigskip\bigskip\bigskip
\vbox{\vskip -.1in\hbox{\centerline{\epsffile{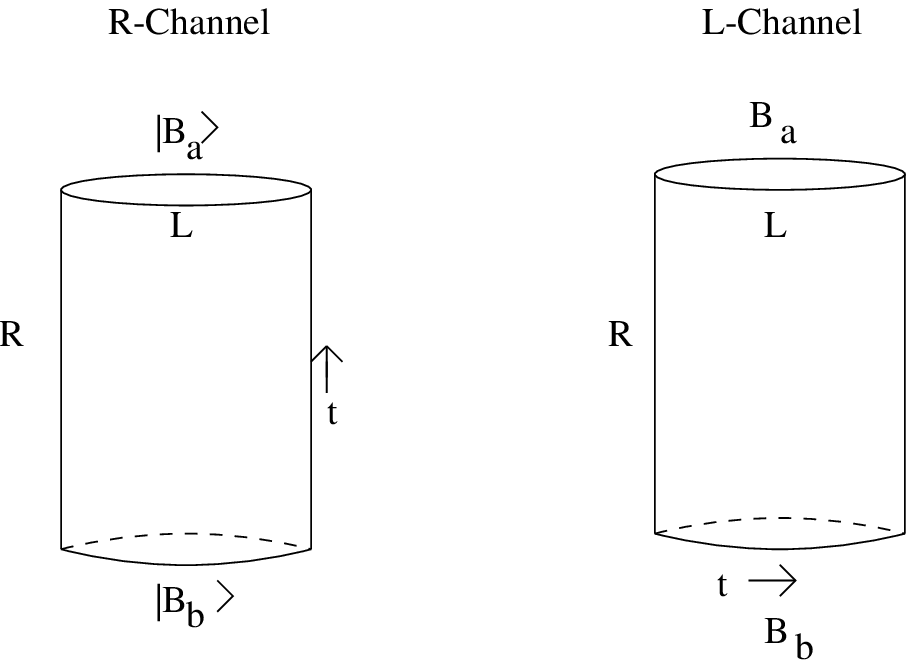}}}
\vskip .1in
{\leftskip .5in \rightskip .5in \noindent \ninerm \baselineskip=10pt
Figure 1.
Cylindrical geometry in the R and L channels.
\smallskip}} \bigskip
\endinsert

In the second part of this paper we address the same questions for a non
scalar theory, namely the sine-Gordon model with Dirichlet
boundary conditions. This case is technically
more difficult. Instead of rederiving a TBA in the R channel, we have
generalized the beautiful approach of Destri and de Vega \DDV\ for a system
with
boundaries. This requires working for a while with the lattice theory, here
chosen to be the XXZ model with boundary magnetic fields.

In section 6 we recall the various elements of the model, and carry out the
TBA in the L channel. As a result, we are able to obtain the boundary ground
state energy of the sine-Gordon model with Dirichlet boundary conditions. In
section 7 we write the ground state energy. In section 8 we use this result
for minimal models, proposing in particular boundary conditions that
correspond to excited states of the theory. Section 9 collects conclusions.
In particular we show that the DDV equations are actually essentially similar
to the equations for a scalar theory derived in section 3.


\vfill\eject

\centerline{\bf Part I: Scalar theories}

\bigskip

\newsec{Ground state energy of the Ising model with boundaries}

In the bulk, the Ising model can be described in terms of massive fermionic
operators $A(\t)$ and $A^{\dagger}(\t)$ with the usual anti-commutation
relations, $\{A(\t),A^{\dagger}(\t')\} = 2\pi \delta(\t-\t')$,
where $\t$ is the rapidity variable. The mass $m$ of the fermionic field
is proportional to  $T_c - T$. The $S$-matrix in the bulk is simply
$-1$. In this section, we are interested in the explicit computation of the
ground state energy of this model for different boundary conditions. Let us
consider then the Ising model in the low temperature phase ($T < T_c$) in the
geometry of figure 2, i.e. a long strip of horizontal length $L$ and vertical
width $R$. At the extremities of the horizontal axis, the order parameter of
the model is subjected to periodic boundary conditions while in the vertical
direction it satisfies boundary conditions of type $a$ and $b$. Thus, we are
effectively considering the partition function on a cylinder of circumference
$L$ and length $R$, with different boundary conditions on each end of the
cylinder. (See figure 1.)
As we discuss in more detail below, such boundary conditions are
expressed in terms of boundary states $\mid B_a >$ and $\mid B_b>$.

\midinsert
\epsfxsize = 3in
\bigskip\bigskip\bigskip\bigskip
\vbox{\vskip -.1in\hbox{\centerline{\epsffile{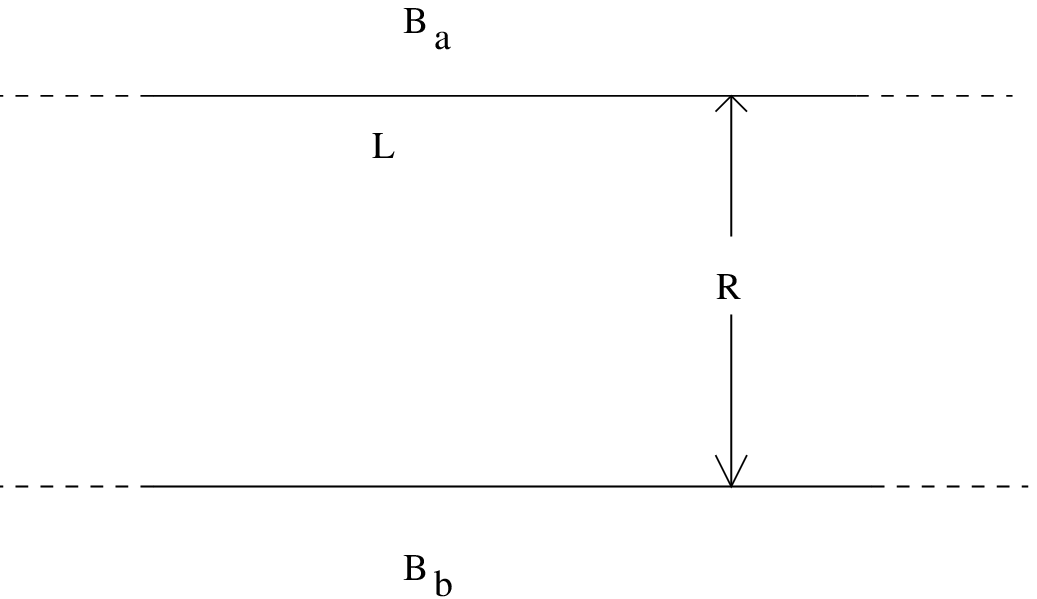}}}
\vskip .1in
{\leftskip .5in \rightskip .5in \noindent \ninerm \baselineskip=10pt
Figure 2.
Strip geometry.
\smallskip}} \bigskip
\endinsert

Depending of our quantization scheme, we have two possible ways to
compute the partition function of the system. The first possibility consists
in choosing as direction of time the horizontal axis and therefore the
partition function will be expressed as
\eqn\part
{
Z_{ab} = Tr \, e^{-L H_{ab}}\, ,
}
where $H_{ab}$ is the Hamiltonian relative to the system with boundary
conditions $(a,b)$. In the second method the time evolution takes place
along the vertical axis and therefore the partition function is given by
the matrix element of the time evolution operator between the boundary states,
i.e.
\eqn\partt{
Z_{ab} = <B_a\mid e^{-R H}\mid B_b>\, ,
}
where now $H$ is the Hamiltonian of the bulk system. The ground state energy
$E_{ab}(R)$ is, by definition, the leading term arising in the large $L$ limit
of the first expression, eq. \part, i.e.
\eqn\Llim{
Z_{ab} \sim e^{-L E_{ab}^0(R)}\,.}
However, in view of the equivalence of the two quantization schemes, we can
compute this quantity by looking at the large $L$ limit
of the second expression, eq. \partt\ .

Since eq. \partt\ employs the boundary states of the model, let us shortly
recall their basic properties (for more detail see the original
reference \GZ). The Ising spins placed on the boundaries can be subjected
to three possible boundary conditions, namely: (i) they can be frozen to one
of the two possible fixed values $\pm$  (fixed boundary condition); (ii)
they can be completely free to fluctuate (free boundary condition); (iii)
they can be coupled to a boundary magnetic field $h$ (magnetic boundary
condition). In the QFT description of the model, each of the above microscopic
configurations corresponds to a boundary state $\mid B>$, which for infinite
length $L$, reads \GZ
\eqn\bounk{
\mid B> = g\exp\left[\int_0^{\infty} {d\t \over 2 \pi} K(\t)
A^{\dagger}(-\t) A^{\dagger}(\t)\right]\mid 0> .}
Here $g$ is an overall normalization which we shall discuss later. Since we
are for the moment interested in the large $L$ limit, we simply set this
factor equal to unity. For simplicity, we ignore possible additional
contributions to the boundary state from zero momentum particles. From the
point of view of QFT, the boundary state can be therefore regarded as a
particular state of the Hilbert space of the bulk theory, made of a
superposition of pairs of particles of equal and opposite momentum
(``Cooper pairs''). All information relative to a particular boundary
condition is encoded into the function $K(\t)$ which can be seen as the
elementary amplitude to create a virtual pair of particles.

For the Ising model at low temperature, the vacuum state $\mid 0>$ can be one
of the two possible vacua of the model in this phase and the explicit
expressions of the amplitude $K(\t)$ for the three cases above considered
are given by \GZ
\eqn\free{
K_{free} = -i \coth{\t\over 2}\, ,}
for the free boundary conditions,
\eqn\fixed{
K_{fixed} = i \tanh{\t\over 2}\, ,}
for the fixed boundary conditions, and
\eqn\magn{
K_{h} = i \tanh{\t\over 2} {{k + \cosh\t}\over {k -\cosh\t}}\, ,}
for magnetic boundary conditions, where $k= 1 - {h^2\over 2 m}$. It is
evident that by varying $h$, we can interpolate between the free boundary
condition ($h = 0$) and the fixed one ($h \rightarrow \infty$).

After this brief discussion on the boundary states, let us come back now to
the evaluation of eq. \partt\ in the limit $L\rightarrow \infty$. The first
thing to consider is the action of the time evolution operator $e^{-R H}$
on the boundary state $\mid B_b>$. Since $H$ is the Hamiltonian of the
bulk theory, it is diagonalized on the basis of the multiparticle asymptotic
states $\mid \t_1,\ldots,\t_n>$,  with eigenvalues given by
$m\sum_i^n \cosh\t_i$. Expanding the boundary state $\mid B_b>$ in terms of
multiparticle states by using its definition \bounk, it is easy to see that
the action of the time evolution operator on $\mid B_b>$ simply reduces to
a redefinition of the amplitude $K_b(\t)$ of this state, namely
\eqn\red{
K_b(\t) \rightarrow K_b(\t,R) \equiv K_b(\t) e^{-2 mR \cosh\t}\, .}
Consequently, from an abstract point of view, the evaluation of the partition
function $Z_{ab}$ consists in computing the scalar product of two boundary
states, one of them $R$ dependent, i.e. $Z_{ab} = <B_a \mid B_b (R)>$.
To determine such  a quantity, it is convenient to expand both boundary
states in their multiparticle components. For the integrability of the theory,
transition amplitudes between states with different number of particles are
not allowed and therefore $Z_{ab}$ can be expressed as
\eqn\mult{
Z_{ab} = \sum_{N=0}^{\infty} {\cal Z}_{ab}^{(2N)}\, ,}
where
\eqn\epartial{\eqalign{
{\cal Z}_{ab}^{(2N)} & = {1\over (N!)^2} \int_0^{\infty}
\prod_{i,j=1}^N {d\t_i\over 2 \pi} {d\theta'_j\over 2 \pi}
\bar K_a(\t_i) K_b(\theta'_j,R) \cr
& <0\mid A(-\t'_1) A(\t'_1) \ldots A(-\t'_N) A(\t'_N)
A^{\dagger}(\theta_N) A^{\dagger}(-\theta_N) \ldots
A^{\dagger}(\theta_1) A^{\dagger}(-\theta_1) \mid 0>\cr}\, .}

The first term of this series is given by $Z_{ab}^{(0)}=<0\mid 0>=1$. The
higher
terms can be computed by using the Wick theorem. There is, however, a subtlety
which already arises in the computation of the second term
\eqn\sec{\eqalign{
{\cal Z}_{ab}^{(2)} = & \int_{0}^{\infty} {d\t \over 2 \pi}
\int_{0}^{\infty} {d\theta' \over 2 \pi} \bar K_a(\t) K_b(\theta',R)
(2 \pi)^2 \delta(\t-\theta') \delta(\t-\theta') = \cr
& = \delta(0) \int_0^{\infty} {d\t\over 2\pi} \bar K_a(\t) K_b(\t)
e^{-2 m R \cosh\t} \cr}\, .}
The term $\delta(0)$ signals the divergence due to the infinite volume limit
on which the boundary states are defined. To correctly extract the
$L$ dependence of this expression, we need to redefine the normalization of
the multiparticle states, i.e. to provide a regularization of the $\delta$
function. If we were working in the momentum space, a possible regularization
of the $\delta$ function is given by
\eqn\delreg{
\delta_L(p) = {1\over 2\pi} \int_{-{L\over 2}}^{{L\over 2}} dx \, e^{i p x} =
{1\over \pi p} \sin{p L \over 2}\, .}
Using this definition, we can make sense of the square of the $\delta$ function
and in the limit $L\rightarrow \infty$ we have $(\delta_L(p-q))^2
\rightarrow {L\over 2\pi} \delta(p-q)$. However, working in the rapidity
variable, the final expression of the square of the $\delta$ function
entering \sec\ acquires an extra term $m\cosh\t$ due to the Jacobian of the
transformation from the momentum to rapidity variable and the final
result is to substitute
\eqn\esub{[\delta(\t-\theta')]^2 \rightarrow {mL\over 2\pi}
\cosh\t \,\delta(\t - \theta')\, .}
This regularization allows us to keep track of the $L$ dependence of our
expressions and therefore to extract the ground state energy $E_{a,b}^0$.
With this substitution, $Z_{ab}^{(2)}$ is now
given by
\eqn\two{Z_{ab}^{(2)} = m L \int_{0}^{\infty} {d\t \over 2\pi} \cosh\t
\bar K_a(\t) K_b(\t) e^{-2 m R \cosh\t}\, .}
For later convenience, it is useful to define the following quantities
\eqn\ik{I_N = \int_0^{\infty} {d\t\over 2\pi} \cosh\t
\left[\,\,\bar K_a(\t) K_b(\t) e^{-2 m R \cosh\t}\,\right]^N\, .}
With the new normalization of the states, the higher terms
${\cal Z}_{a,b}^{(2N)}$ in the series \mult\ are polynomials in the variable
$(mL)$ of order $N$, as explicitly shown by the first representatives
\eqn\nex{\eqalign{
& Z_{ab}^{(4)} = {(mL)^2\over 2} I_1^2 - {(mL)\over 2} I_2 \, ,\cr
& Z_{ab}^{(6)} = {(mL)^3\over 3!} I_1^3 - {(mL)^2 \over 2} I_1 I_2 + {(mL)\over
3}
I_3 \, ,\cr
& Z_{ab}^{(8)} = {(mL)^4\over 4!} I_1^4 - (mL)^3 I_1^2 I_2 + {(mL)^2\over 2}
\left[{2\over 3} I_1 I_3 + \left({I_2\over 2}\right)^2\right] -
{(mL)\over 4} I_4 \, ,\cr
}}
and so on. The resulting sum organizes into an exponential series
\eqn\esp{
Z_{ab} =  1 + (mL) f(R) + {(mL)^2\over 2!} (f(R))^2 +
\ldots {(mL)^n \over n!} (f(R))^n + \ldots = \exp[mL f(R)] }
where
\eqn\f{
f(R) = \sum_{N=1}^{\infty} (-1)^{N+1} {I_N\over N} =
\int_0^{\infty} {d\t\over 2\pi} \cosh\t \log\left(1 + \bar K_a(\t)
K_b(\t) e^{-2 m R \cosh\t}\right)\, .}
Therefore, the ground state energy for boundary conditions $(a,b)$ is given by
\eqn\ge{
E_{ab}^0(R) = - {m\over 4\pi}\int_{-\infty}^{\infty}
d\t \cosh\t \log\left(1 + \bar K_a(\t)
K_b(\t) e^{-2 m R \cosh\t}\right)\, .}

The above result can also be easily derived
as a special case of the general
formula
\eqn\edet{
<0| e^{( \tilde{a} , M a ) } e^{(a^\dagger , N \tilde{a}^\dagger )} |0>
= \det (1+ NM) ,}
where
$(\tilde{a} , M a) = \sum_{n,m} \tilde{a}_n M_{nm} a_m $ and
$\{ a_n , a^\dagger_m \} = \{ \tilde{a}_n , \tilde{a}^\dagger_m \}
= \delta_{n,m} $, $\{ \tilde{a}_n , a^\dagger_m \} = 0$.  Specializing
to our situation, the matrices $M,N$ are kernels:
$M(\t , \t' ) = \delta (\t - \t' ) \bar{K} (\t ) $,
$N(\t , \t' ) = \delta (\t - \t' ) K (\t )
e^{-2mR \cosh \t } $,
and the determinant in \edet\ is a Fredholm determinant.  Using
$\log \det (1+NM) = tr \log (1+NM)$, and regulating the squares of
delta functions as in \esub, one obtains \ge.

As a check of this expression, we can compare its ultraviolet limit
$R\rightarrow 0$ with the result directly obtained by using CFT \Cardybc .
In the conformal limit, general expressions for the partition function
in the cylindrical geometry with boundaries were derived by Cardy \rcardy.
In the R-channel, one has
\eqn\Rchan{
Z_{ab} = \sum_i n^i_{ab} ~ \chi_i (q) ,}
where  $q= \exp (-\pi L/R )$, $\chi_i$ are the characters,
$\chi_i (q) = q^{-c/24} \,Tr \,q^{L_0}$, and $n^i_{ab}$ denotes how
many times the representation $[i]$ appears with the boundary
conditions $(a,b)$. Thus, as $L$ tends to infinity, one has
\eqn\confe{
E^0_{ab} =  {\pi\over R} \( L^{(0)}_0 - {c\over 24} \), }
where $L^{(0)}_0$ is lowest $L_0$ eigenvalue of the states in
the representations $[i]$ for which $n^i_{ab} \neq 0$.
Taking the limit $R\rightarrow 0$ in \ge , the ground state energy
assumes the scaling form
\eqn\sca{
E_{ab}^0(R) \approx - {1\over 4\pi R}
\int_0^{\infty} d\epsilon \log\left[1 + \lambda_{ab}(\infty)
 e^{-\epsilon}\right]\, ,}
where
\eqn\epe{\lambda_{ab}(\theta) = \bar K_a(\t) K_b(\t)\,.
}
Since $\lambda_{free,free}(\infty) = \lambda_{fixed,fixed}(\infty) = 1\, ,$
\eqn\ff{
E_{free,free}^0 = E_{fixed,fixed}^0\approx  -{\pi\over 48R}}
With those boundary conditions, the ground state energy is dictated by
the conformal family of the identity operator. For the mixed boundary
conditions, we have instead $\lambda_{free,fixed}(\infty)=-1 $ and therefore
\eqn\mix{
E_{free,fixed}^0 \approx {\pi\over 24R}\,.}
In this case, the ground state energy is ruled by the conformal family of
the magnetization operator, of conformal dimension $1\over 16$. These results
are in agreement with those obtained by Cardy \Cardybc.

\newsec{The ground state energy of a scalar theory with boundaries and TBA in
the R-channel}

Reproducing the above computation in the case of a scalar interacting theory
is more difficult. This is because the particles scatter non trivially, so
care must be exercised in computing scalar products. Also, the allowed
rapidities of the physical states are non trivial solutions of Bethe type
equations. The simplest way to proceed is to use a ``thermal approach'',
generalizing the standard Thermodynamic Bethe Ansatz (TBA).

The TBA equations in a torus geometry with no boundary were studied
in \rtba. In the cylindrical geometry considered here with boundary
interactions, the two possible ways of viewing the partition function
which correspond to the two possible channels for time evolution are rather
asymmetrical, thus one expects the TBA equations for each channel to take
different forms.

In the analysis of the previous section, we examined the large $L$ behaviour
of the partition function using the boundary states. This corresponds to
viewing the partition function in the `$R$-channel', where time evolution is
in the direction of $R$. As we now describe, one can obtain the TBA equations
in this channel in the limit of large $L$.

In the situation with non-trivial S-matrix, one may attempt to evaluate
\partt\ using the form of the boundary states
\bounk\ and the Faddeev-Zamolodchikov algebra:
\eqn\fatzam{\eqalign{
A^\dag (\t) A^\dag (\t') & = S(\t-\t') A^\dag (\t') A^\dag (\t) \, ,\cr
A(\t) A(\t') & = S(\t-\t') A(\t') A(\t) \, ,\cr
A(\t) A^\dag (\t') & = S(\t-\t') A^\dag (\t') A(\t) + \delta(\t-\t')\, .\cr}}
As in the Ising case discussed in the previous section, one finds
that the result is ill-defined due to the fact that the
formula \bounk\ for boundary states is meant in the limit $L\to\infty$, and
this leads to squares of $\delta$-functions.  In the Ising case, the
regularization \esub\ is adequate since the particles are free and the
quantization condition on the momenta does not involve the S-matrix. More
generally, one can view the following TBA derivation as a way of regularizing
the inner product \partt\ for very large but finite $L$.

For simplicity, consider a  model with a single particle excitation.
Our starting point is eq. \partt\ which can be written as
\eqn\neqi{Z_{ab}=\sum_\alpha {<B_a|\alpha><\alpha|B_b>\over
<\alpha|\alpha>}e^{-RE_\alpha},}
where $|B_a>$ and $|B_b>$ are boundary states and the sum over $\alpha$
stands for a formal sum over all the states of the theory. Due to the form
of the boundary states, the states $|\al >$ that contribute to the sum are
of the form
\eqn\alphrep{\eqalign{
|\alpha_{2N}>&=|\t_N,-\t_N,\ldots,\t_1,-\t_1>\cr
&=A^\dagger (\t_N)A^\dagger (-\t_N)\ldots A^\dagger
(\t_1)A^\dagger (-\t_1)|0>,\cr}}
where
$\t_N>\t_{N-1}>\ldots>\t_1 > 0$ (inequalities are strict, since due
to the condition $S(0) = -1$, the Bethe wave function for particles on a
circle of radius $L$ vanishes when two rapidites are equal).
Using the Faddeev-Zamolodchikov algebra, one finds that
\eqn\eIIIiii{
<B_a | 2N> =  ( \delta (0) )^N \prod_{i=1}^N \bar{K}_a (\th_i )\, ,}
where the formal expression  $(\delta (0))^N$ results
from terms  $\prod_i  \delta (\th_i  -\th_i  )$.
It is simple to show \eIIIiii\ by introducing pair creation operators
$B(\th ) = A(-\th ) A (\th ) $, $B^\dagger (\th ) = A^\dagger (\th )
A^\dagger (-\th ) $, satisfying the algebra
\eqn\eIIIiv{\eqalign{
B(\th ) B(\th ' ) &=  B(\th') B(\th ) \,\,, ~~~~~~~
B^\dagger (\th ) B^\dagger (\th' ) = B^\dagger (\th' ) B^\dagger (\th ) \, ,\cr
B(\th ) B^\dagger (\th') & = B^\dagger (\th') B (\th )
+ \delta^2 (\th - \th' ) \cr
&~~~~
+ \delta(\th -\th') S(\th' - \th ) A^\dagger (-\th') A(-\th )
+ \delta(\th -\th') S(\th - \th' ) A^\dagger (\th') A(\th ) \,. \cr}}
One also finds $<2N|2N> = (\delta (0))^{2N}$, thus,
\eqn\eIIIv{
{ <B_a |2N> <2N| B_b > \over <2N|2N>}
= \prod_{i=1}^N  \bar{K}_a (\th_i ) K_b (\th_i ) \, . }
Let us introduce a density per unit length of pairs of particles
$P(\th )$, so that in \neqi, $E_\al = \sum_{i=1}^N 2m \cosh \th_i $
is replaced with  $2mL \int d\th   \cosh (\th ) P (\th ) $.
Due to the factorization \eIIIv, the partition function reads
\eqn\lnpart{ Z\approx \int [dP]\exp\left\{L\int_0^\infty
\left[ \log (\bar{K}_a(\t)  K_b(\t) ) - 2Rm\cosh\t\right]P(\t)d\t
+ S([P])\right\},}
where $S[P]$ is the entropy of the particle configuration described by
the distribution $P(\th)$. From this equation, we see that the contribution
of the boundaries has a natural interpretation as (rapidity dependent)
chemical potentials.

The density $P(\th )$ is constrained by the quantization condition.
For the states $|2N>$, this reads
\eqn\eother{
e^{imL\sinh (\th_i ) }
S(2\th_i) \prod_{j\neq i} S(\th_i - \th_j ) S(\th_i + \th_j ) = 1 \,. }
Introducing densities of particles and holes as usual
such that $L [P(\th)+P^h (\th )] d\t$ is the number of allowed rapidities
between $\t$ and $\t+d\t$, one has the quantization conditions
\eqn\quant{
2\pi \left(P(\t) + P^h(\t)\right) =
m \cosh\t -2\pi \int_0^{\infty} \left[\Phi(\t - \t') +
\Phi(\t + \t')\right] P(\t') d\t'\, ,}
where $\Phi$ is defined by
\eqn\defff{\Phi(\theta)=-{1\over2i\pi}{d\over d\t}\ln S(\t)\,.}
In terms of $P, P^h$, the entropy takes the usual form:
\eqn\entroo{S([P])=L\int_0^\infty d\t \left[
(P + P^h )\ln(P + P^h) - P \ln P - P^h \ln P^h\right]\,.}
Finally, the TBA equations arise as a saddle point evaluation of
\lnpart , i.e. the minimization of the free energy, subject to the constraint
\quant. Using standard manipulations, one obtains the leading behaviour of
$\ln Z$ for large $L$:
\eqn\ennn{\log Z_{ab} \approx - L E_{ab}^0 (R)  =
{mL\over 2\pi}\int_0^\infty  \cosh\t\ln\left[1 +
e^{-\epsilon(\t)}\right]d\t\, .}
where $\ep$ satisfies
\eqn\tbaeqi{\epsilon(\t) = 2Rm\cosh\t - \log (\bar{K}_a K_b )
+ \Phi_s *  \log (1 + e^{-\ep } ) \,. }
In the previous equation, $\Phi_s \equiv \Phi(\th - \th') + \Phi(\th + \th')$
and $*$ denotes the convolution. The combination $\bar {K}_a (\th ) K_b (\th )$
is an even function of $\th$, since  $\bar{K} (\th ) = K(-\th )$. Thus,
we can extend the domain of definition of $\epsilon(\t)$ to the whole
real axis by $\epsilon(-\t)=\epsilon(\t)$ and let the integrals run from
$-\infty$ to $+\infty$. The equations for the ground state energy
may be recast into the following:
\eqn\finalsystem{\eqalign{
E_{ab}^0(R) = & -{m\over 4\pi}\int_{-\infty}^\infty
\cosh\t\log\left[1 +
\lambda_{ab}(\t) e^{-\epsilon(\t)}\right]d\t\, ,\cr
\epsilon(\t) = & 2Rm\cosh\t + \int_{-\infty}^\infty
\Phi(\t-\t')\log \left[1 +
\lambda_{ab}(\t') e^{-\epsilon(\t')}\right]d\t',\cr}}
where
$$\lambda_{ab} (\th ) = \bar{K}_a (\th ) K_b (\th ) \,.$$

One may now see explicitly how the TBA regularized the inner product
\partt. A multiparticle expansion of the TBA equations \ennn\tbaeqi\
can be obtained by solving them iteratively as follows:
\eqn\eiter{
\ep_{n+1} = \ep_0 + \Phi_s * \log (1+ e^{-\ep_n} ) \,,}
starting from $\ep_0 = 2mR\cosh \th  - \log (\bar{K}_a K_b ) $.
One obtains
\eqn\emult{
Z_{ab} = 1 + {mL\over 2\pi} (D_1 + D_2 +\ldots) + \inv{2} \( {mL\over 2\pi}
\)^2 D_1^2 +\ldots, }
where
\eqn\eds{\eqalign{
D_1 &=  \int_{-\infty}^\infty d\th \cosh \th e^{-2mR \cosh \th } \bar{K}_a
(\th)  K_b (\th )
\, ,
\cr
D_2 &=  -\inv{2} \int_{-\infty}^\infty d\th \cosh \th e^{-4mR\cosh \th }
\bar{K}^2_a (\th ) K^2_b (\th ) + \inv{2\pi} \int_{-\infty}^\infty d\th d\th'
e^{-2mR(\cosh \th + \cosh \th' )} \cr
& ~~~~
\( \Phi (\th + \th') + \Phi (\th - \th') \)
\bar{K}_a (\th') K_b (\th') \bar{K}_a (\th ) K_b (\th ) . \cr}}
On the other hand, evaluating \partt\ using the Faddeev-Zamolodchikov
algebra, one obtains
\eqn\emultb{\eqalign{
Z_{ab} &= 1  + \int d\th \delta (0) \{ \bar{K}_a  K_b
e^{-2mR \cosh \th }  -\inv{2} \bar{K}^2_a K^2_b  e^{-4mR \cosh \th } \}
+ \cr
&~~~~~ + \inv{2} \int d\th d\th' \delta^2 (0)
\bar{K}_a (\th ) K_b (\th ) \bar{K}_a (\th ')  K_b (\th' )
e^{-2mR(\cosh \th + \cosh \th' )}  +.... \cr }}
Comparing \emult\ with \emultb, one sees that the single $\delta$-functions
in \emultb\ were regulated as in \esub, however the $\delta^2 (0) $ is
regulated in a more complicated fashion involving the S-matrix:
\eqn\ereg{
\delta(\th - \th ) \delta (\th' -\th' )
\to \({mL\over 2\pi} \)^2 \cosh \th \cosh \th' + {mL\over 2\pi^2}
\cosh \th \( \Phi (\th + \th') + \Phi (\th -\th' ) \) . }

As an explicit example of our discussion, we will consider the $\varphi_{1,3}$
massive deformation of the minimal non-unitarity conformal model
${\cal M}_{3,5}$, in the R-channel. The central charge is $c = -3/5$ and
there are four primary fields of chiral dimensions
$(0,-{1\over 20},{1\over 5},{3\over 4})$. The massive phase of this
deformation has only one massive particle with the two-particle elastic
$S$-matrix given by $S(\t) = - i \tanh\half
\left(\t - i {\pi\over 2}\right)$ \ressmir . For the boundary reflection
amplitudes, solutions of the equations
\eqn\teq{\eqalign{
& {\cal R}(\t) {\cal R}(-\t) = 1 \, ,\cr
& {\cal R}\left({i\pi\over 2} -\t\right) = S(2\t) {\cal R}\left({i\pi\over 2} +
\t\right)\, ,\cr}}
we have two minimal solutions, given by
\eqn\solu{\eqalign{
& {\cal R}_1(\t) = -i \tanh\left({\t\over 2} - {i\pi\over 4}\right)
{\sinh\left({\t\over 2} - {i\pi\over 8}\right)\over
\sinh\left({\t\over 2} + {i\pi\over 8}\right)} \, ,\cr
& {\cal R}_2(\t) = i \coth\left({\t\over 2} - {i\pi\over 4}\right)
{\sinh\left({\t\over 2} - {i\pi\over 8}\right)\over
\sinh\left({\t\over 2} + {i\pi\over 8}\right)}\, ,\cr}}
and from those, we can compute
$K(\t) = {\cal R}\left({i\pi\over 2} - \t\right)$. We now want to extract
the ultraviolet limit of the ground state energy for different boundary
conditions associated to these amplitudes. The kernel entering the TBA
equation is given by $\Phi(\t) = 1/\cosh\t$. As in analogous TBA computation,
the ultraviolet limit of the ground state energy is ruled by the "kink"
solution of the pseudoenergy $\epsilon(\t)$, i.e. the universal function
which flattens in the central region $-\log(2/r) \ll \t \ll \log(2/r)$
to the constant value $\epsilon_{ab}^0$, solution in this case of the
algebraic equation
\eqn\const{
\epsilon_{ab}^0 = -\half \log \left[1 + \lambda_{ab}(\infty)
e^{-\epsilon_{ab}^0}
\right]\, .}
In fact, in the ultraviolet limit $mR \rightarrow 0$, we have
\eqn\enuv{
E_{ab}^{0} \approx -{1\over 4 \pi R}\int_{\epsilon_{ab}^0}^{\infty}
d\epsilon \left\{{\epsilon \lambda_{ab}(\infty) e^{-\epsilon}
\over 1 + \lambda_{ab}(\infty) e^{-\epsilon}} +
\log\left[1 + \lambda_{ab}(\infty) e^{-\epsilon}\right]\right\}\, .}
Choosing boundary conditions of the type $(1,1)$ or $(2,2)$ on both sides
of the strip, we have $\lambda_{11}(\infty) = \lambda_{22}(\infty) = 1$ and
for the constant value of the pseudoenergy we have
\eqn\pse{
\epsilon_{11}^0 = \epsilon_{22}^{0} = - \log\left({1 + {\sqrt 5}\over
2}\right)\, ,}
and as a final result
\eqn\final{
E_{11}^0 = E_{22}^0 \approx -{\pi\over 40 R}\, .}
On the other hand, choosing boundary conditions of the type $(1,2)$, we have
$\lambda_{12}(\infty) = -1$. In this situation we have
\eqn\ps{
\epsilon_{12}^0 = -\log\left({-1 + {\sqrt 5}\over 2}\right)\, ,}
and correspondingly
\eqn\fin{
E_{12}^0 (R) \approx  {\pi\over 40 R}\,.}
In terms of CFT, in the first two cases the boundary conditions select
as dominant contribution in the partition function the one of the
conformal chiral field $\sigma$ of anomalous dimension $\Delta = -{1\over 20}$
whereas in the last case, the leading contribution comes from the identity
family.

In closing this section, observe that by taking the limit $R\to 0$ in the
above equations \ennn\tbaeqi , we obtain the leading behaviour for the
scalar product of boundary states
\eqn\scalprod{\eqalign{\ln<B_a|B_b>\approx&\quad {mL\over
4\pi}\int_{-\infty}^\infty
\cosh\theta\ln\left[1+\lambda_{ab}e^{-\epsilon(\t)}\right]d\t\, ,\cr
\epsilon(\t)=&\int_{-\infty}^\infty
\Phi(\t-\t')\ln\left[1+\lambda_{ab}(\t)e^{-\epsilon(\t)}\right]d\t\,.\cr}}

\newsec{Normalizations and TBA in the L-channel}

We can also discuss in more detail the overall normalization alluded to in
\bounk. There, $|0>$ is the ground state of the massive theory, which in the
renormalized theory is simply a state without any excitation. Although the
correspondence between particle states of the massive theory and conformal
states in the UV limit is not totally known, the ground states usually match,
that is in the UV limit, $|0>$ in \bounk\  goes to the conformal ground state
$|0>_c$, while all the states with particles go to combination of fields and
descendants with higher conformal weights. Therefore, if the boundary state
\bounk\ corresponds to some state $|B>_c$ in the conformal theory,
we expect
\eqn\eqqqq{ g = <0|B(m)> = \, _c<0|B>_c \hbox{ for any } m .}
Let us further investigate this identity. In the ``L-channel'', time evolution
is along the circumference of the cylinder of length $L$, and the Hilbert
space is constructed along the finite segment of length $R$. In the limit of
large $R$, TBA equations for the partition function were obtained in \FS.
In this channel, the particles are confined to the segment of length $R$ and
reflect off the boundary with reflection amplitude ${\cal R}_a (\theta )$ at
one boundary, and ${\cal R}_b (\theta )$ at the other boundary. The
quantization condition on the momenta involves in this case the boundary
reflection amplitudes:
\eqn\equan{
e^{2imR \sinh (\th_i ) }
\[ \prod_{j\neq i} S(\th_i - \th_j ) S(\th_i + \th_j ) \]
{\cal R}_a (\th_i ) {\cal R}_b (\th_i ) = 1 \, , }
where  $\t_i>0$. Introduce the density of particles and holes as usual, so
$2 R [P(\t)+P^h(\t)] d\t$ is the number of allowed rapidities between $\t$
and $\t+d\t$. It is also convenient to let rapidities run from $-\infty$ to
$\infty$, defining the densities for negative rapidities by parity, so
\equan\ reads
\eqn\quantnew{2\pi(P(\t) + P^h(\t)) = m\cosh\t -
2\pi\int_{-\infty}^\infty\Phi(\t-\t')
P(\t') + {\Theta_{ab}(\t)\over 2R}\, ,}
where $\Phi$ is as before, and
\eqn\eth{
\Th_{ab}  (\th ) =
{1\over i}{d\over d\th} \log ({\cal R}_a {\cal R}_b )
-{1\over i} {d\over d\th}
 \log S(2\th ) - 2\pi \delta
(\th ) \, , }
and we used the fact that, by unitarity, $\Th_{ab}$ is an even function.
In the latter expression, the $\delta$ term arises because
vanishing rapidities, although possible solutions of \equan, are not
acceptable since the corresponding wave function vanishes exactly.

{}From here the standard TBA procedure leads to the following result:
\eqn\eLchan{\eqalign{
\log Z_{ab} \approx&  {1\over 4\pi}\int_{-\infty}^\infty
  \left[ 2m R \cosh \th + \Th_{ab}
(\th ) \right]
\log \left[1+ e^{-\ep (\th )} \right] d\th \, ,\cr
\ep (\th ) = &mL \cosh \t   +  \int_{-\infty}^\infty \Phi(\t-\t')
 \log \left[1+ e^{-\ep(\t')} \right]d\t'\, .\cr} }
The $\Th $ term in \eLchan\  is $R$ independent and may be considered as a
boundary free energy term. As discussed in \FS\ the TBA actually reproduces
the boundary free energy up to an additive constant. This is because, when
considering contributions of order one, one should take care of various
corrections like corrections to the Stirling formula in evaluating the
entropy, or loop corrections to the saddle point, which we have discarded
here. Also the correspondence between the entropy of the field theory and the
one computed using particles might involve some constant, for instance
if the particles describe kinks and an overall choice of configuration
is left out.

Now we can interpret this boundary free energy easily. Indeed in the large
$R$ limit, the next to leading behaviour of the partition function
\partt\ is
\eqn\nexttol{Z_{ab}\approx <B_a |0><0|B_b > e^{-R  E_0(R)},}
up to exponentially small terms, where $E_0$ is the ground state
energy of periodic Hamiltonian. Therefore we have
\eqn\gcompp{\ln <B_a |0><0|B_b > = {1\over 4\pi}\int_{-\infty}^\infty
 \Theta_{ab}(\t) \ln\left[1+e^{-\epsilon(\t)}\right]d\t
+\hbox{cst}.}
In the IR limit given by $L \to \infty$, $\epsilon$ goes to
infinity and the first term
on the rhs of \gcompp\ vanishes. In the UV limit, expressed by
$L \to 0$, on the contrary $\epsilon$ goes to a constant. The statement
\eqqqq\ translates therefore into the condition
\eqn\condo{\int_{-\infty}^\infty \Theta(\t)d\t = 0 \, .}
This condition is indeed true, as can be easily seen by using \eth\ and
the cross unitarity relation
\eqn\crossunnn{{\cal R}\left(i{\pi\over 2}-\t\right) = S(2\t)
{\cal R} \left(i{\pi\over 2}+\t\right),}
together with analyticity.

As we will see in the next section, for finite $L$, $g$ can be
a non-trivial function of $L$.

\newsec{Massless Theory in the Bulk}

An interesting situation is when the bulk theory is massless, but the
conformal invariance is broken by the boundary interactions. The previous
formalism carries over easily to that case. This time the bulk
theory is described by massless particles which are left or right moving
with trivial left-right scattering. Parametrize the right (left) moving
particles by rapidities so that $E=p={m\over 2}e^\t$ (resp.
$E=-p={m\over 2}e^{-\t}$) (observe that opposite momenta still correspond
to opposite rapidities). Then the left-left and right-right scattering
are described by an identical $S$ matrix in terms of rapidities, and one can
introduce as before
\eqn\masslessphase{\Phi(\t)=-{1\over 2i\pi}{d\over d\t}\ln S_{LL}(\t).}
Now all rapidities in $[-\infty,\infty]$ are allowed, so the quantization
condition reads
\eqn\equivvvv{2\pi(P(\t) + P^h(\t)) =
{m\over 2}e^{\t}-2\pi\int_{-\infty}^\infty
\Phi(\t-\t') P(\th') d\t',}
and for the entropy \entroo\ the integration runs from $-\infty$ to $\infty$.
The energy reads $E=\sum_{i=1}^N me^{\t_i}$. To have a breaking of the
conformal invariance and a corresponding flow, one needs to put an energy
scale at the boundary, so the boundary $S$ matrix depends on some extra
parameter (e.g. a magnetic field), corresponding to an amplitude
$K_a(\t-\t_{B_a})$. The boundary state is then a combination of massless states
involving pairs of left and right moving particles with opposite momenta
(hence opposite rapidites using the above parametrization)
\eqn\newbdr{|B_a>=g\exp\left[\int_{-\infty}^\infty {d\t\over
2\pi} K_a(\t -\t_{B_a}) A^\dagger_L(-\t)A^\dagger_R(\t)\right]|0>.}
Then \eIIIv\  still holds. In the ``R-channel'', the maximization of the free
energy leads now to
\eqn\newsystem{\eqalign{E_{ab}^0&=-{m\over 4\pi}\int_{-\infty}^\infty
e^\t\ln\left[1+\lambda_{ab}(\t)e^{-\epsilon(\t)}\right]d\t\cr
\epsilon(\t)&=mRe^\t+\int_{-\infty}^\infty
\Phi(\t-\t')\ln\left[1+\lambda_{ab}(\t')e^{-\epsilon(\t')}\right]d\t',\cr}}
where
\eqn\lamm{\lambda_{ab}(\t) = \bar{K_a}(\t-\t_{B_a}) K_b(\t-\t_{B_b}).}
It is also possible to compute the boundary entropies by using the
TBA in the ``L-channel'', as above. One finds
\eqn\newbdrent{\ln<B_a |0><0|B_b >
= {1\over 2\pi}\int_{-\infty}^\infty \Theta_{ab}(\t)\ln\left[
1+e^{-\epsilon(\t)}\right] d\t+\hbox{cst},}
where
\eqn\lastnew{\epsilon(\t) = mL{e^\t\over 2}+\int_{-\infty}^\infty
\Phi(\t-\t')\ln \left[1+e^{-\epsilon(\t')}\right]d\t',}
and
\eqn\lastlast{\Theta_{ab}(\t) = {1\over i}{d\over
d\t}\ln\left[{\cal R}_a(\t-\t_{B_a}) {\cal R}_b(\t-\t_{B_b})\right].}
In the last expression compared to \eth\ the term $S(2\t)$ has disappeared
because it would arise from LR scattering which is rapidity independent.
Similarly, vanishing momentum corresponds now to vanishing energy and the
corresponding term can be discarded.

Despite the close similarity of the massless and massive TBA equations,
what is meant by ultraviolet and infrared limit in the massless case is
different from the massive situation. In fact, instead of the bulk mass
parameter, we now vary the boundary energy scale $me^{\theta_B}$ ($m$ just
fixes the overall scale and can be considered constant) which corresponds
physically to varying the boundary interaction. Boundary states in general
will correspond to conformal states both in the UV and in the IR. These
conformal states in general will have different scalar products with the
conformal ground state, and the boundary entropy will vary along the flow.
Accordingly, eq. \condo\ does not hold. Physically, the boundary state
\newbdr\ can now have varying scalar product with the ground state because
conformal states can be represented as combinations of massless particle
states.

To illustrate this situation, consider the Ising model at $T = T_c$ but
with a boundary magnetic field. This induces a flow from the free to the
fixed boundary conditions. The flow is described by a massless boundary matrix
\eqn\massising{{\cal R}(\t-\t_B) = -i
\tanh\left({\t-\t_B\over 2}-{i\pi\over 4}\right).}
Using \newbdrent and setting $g=<0|B>$, one finds
\eqn\gfact{\ln g = \int_{-\infty}^\infty {\ln\left[1+e^{-e^\t}\right]\over
\cosh\left(\t-\ln L/ L_B\right)} \, {d\t\over 2\pi}+\hbox{cst},}
where $L_B = {2\over m}e^{-\t_B}$. This is easily evaluated
\eqn\eval{g=\hbox{cst}{\sqrt{2\pi}\over \Gamma(a+1/2)}\left({a\over
e}\right)^a,}
where $a=L/L^B$. The UV limit is $L\to 0$ and $g\to\hbox{cst} \sqrt{2}$, while
the
IR limit is $L\to \infty$ and $g\to\hbox{cst} 1$. Therefore we find
$g^{UV}/g^{IR} = \sqrt{2} = g^{free}/g^{fixed}$ in agreement with \AL.

The above characterization of the boundary state involving massless
particles is appropriate for very large $L$.  For finite $L$
another characterization is in principle possible which does not
involve massless particles, as we now describe.
Viewing the partition function in the R-channel, the Hamiltonian and the
Hilbert space are the same as for the conformal field theory
on a cylinder.  Therefore, the boundary states can be expressed in terms of
conformal operators even though the boundary states themselves break
the conformal invariance.  Furthermore, since the conformal field
theory is well defined for finite $L$, one can obtain exact partition
functions for both finite $L$ and $R$.  For the Ising case, this was
studied by Chatterjee \rchat \foot{We obtained the results in \rchat\
independently, in a manner presented below.}. For perturbed minimal models,
these boundary states were characterized in \rbaz. In this section, we
show how formulas of the previous sections in part characterize the boundary
states in the simple case of the Ising model.

\def\psib{\bar{\psi}}

Let $\psi (z), \bar{\psi}(\zb )$ denote the left and right components
of the Ising free fermion field, where $z=t+ix$, $\zb = t-ix$. On a cylinder
with $x$ varying along the circumference of length $L$ and $t$ varying along
its length $R$, the mode expansions read
\eqn\emode{\eqalign{
\psi (z) &= \sum_r  \psi_r  \exp \( -2\pi r z/L \)  \cr
\bar{\psi} (\zb) &= \sum_r  \bar{\psi}_r  \exp \( -2\pi r \zb /L \),  \cr
}}
where $r\in \Zmath + 1/2$ ($r\in \Zmath$) in the Neveu-Schwarz (Ramond)
sector.  The interpolating boundary condition takes the form\GZ:
\eqn\eIVii{
{d\over dx} \( \psi - \bar{\psi} \) -i {h^2\over 2} \( \psi + \bar{\psi} \) =0
{}~~~~~~(t=0).}
Inserting the mode expansions \emode\ into \eIVii, one obtains
\eqn\eIViii{
\psib_n  =  a_n \psi_{-n}  , }
where
\eqn\eIViv{
a_n = {n-a\over n+a} , ~~~~~~   a= {h^2 L\over 4\pi}  . }
The equation \eIViii\ implies that the boundary state must
satisfy
\eqn\eIVv{
\( \psib_n -  a_n \psi_{-n} \)  |B> = 0.}
This fixes the boundary states up to a constants $g$, which can depend on
$L$ and $h$:
\eqn\ebound{\eqalign{
|B>^{NS} &=  g_+ \exp \( \sum_{r\in \Zmath + 1/2, r>0}
 a_r  \psib_{-r} \psi_{-r}  \)  |0> \cr
|B>^{R} &=  g_- \exp \( \sum_{n\in \Zmath , n>0}
 a_n  \psib_{-n} \psi_{-n}  \)  |\sigma> , \cr}}
where
$|\sigma>$ is the spin field state of dimension $1/8$ in the
Ramond sector.

The factors $g_\pm$ depend only on $L$ and $h$, and are characterized
by $g_+ = <0|B>^{NS}$, $g_- = <0|B>^R $.  Setting the length $R$
to zero and one of the boundary scattering matrices ${\cal R}_a $ to $1$
in \eLchan, one sees that the $g$-factors correspond to the boundary free
energy term ($\Th$ term). Since the S-matrix is $-1$, the integral equation
just implies $\ep = mL \cosh \th $. Thus $g_\pm$ are just the massless limit
of the expressions:
\eqn\eIVv{
\log g_\pm = \lim_{m\to 0}
\int_0^\infty  {d\th\over 2\pi}
\(-i {d\over d\th} \log R(\th ) - \pi \delta (\th ) \)
\log (1\pm e^{-mL\cosh \th } ).  }
The extra minus sign in $g_-$ versus $g_+$ is easily obtain by including a
$(-1)^F$ in the TBA computation, where $F$ is fermion number.

One has ${\cal R}(\th ) = K({i\pi\over 2} - \th )$, where $K$ is given in
\magn.  Defining $x=mL\cosh \th /2\pi a $ and letting $m\to 0$, one
obtains
\eqn\eIVvi{\eqalign{
\log g_+ &= \inv{\pi} \int_0^\infty \inv{1+x^2}
\log (1 + e^{-2\pi a x } )
\cr
\log g_- &= \inv{\pi} \int_0^\infty \inv{1+x^2}
\log (1 - e^{-2\pi a x } )
+ \inv{4} \log 2  .\cr}}
The extra term in $\log g_-$ arises from the contribution
\eqn\extra{
\lim_{m \to 0} ~  \int_0^\infty \frac{d\th}{2\pi}
\( \inv{\cosh \th }   - \pi \delta(\th ) \)
\log (1 - e^{-mr\cosh \th } )  = \inv{4} \log 2. }
(The $m\to 0$ limit is obtained from
$\log (1 - e^{-mr \cosh \th} ) \approx \log mr + \log \cosh \th $;
the $\log mr$ term vanishes, and the $\log \cosh \th$ term
gives \extra).
The above integrals are easily done\foot{The results for these
integrals presented
in \ref\rgr{I. S. Gradshteyn and I. M. Ryzhik,
{\it Table of Integrals, Series, and Products}, Academic Press.}
are off by a minus sign.}:
\eqn\eIVvii{
g_+ = {\sqrt{2\pi}\over\Gamma (a+1/2 )} \( {a\over e} \)^a ,
{}~~~~g_- = 2^{1/4} {\sqrt{2 \pi a } \over \Gamma (a+1 )}
\( {a\over e} \)^a
.}

The complete boundary state is a linear combination of the NS and R
boundary states:
\eqn\eIVviii{
|B_h> = \inv{\sqrt{2}}
\( |B>^{NS}  + {\rm sign}(h) |B>^R \) . }
The partition function with magnetic fields $h, h'$ on the boundaries is now
easily computed as
\eqn\eIVix{
Z_{h' , h} =  <B_{h'} | e^{-HR} |B_h>  , }
where $H$ is the hamiltonian on the cylinder:
\eqn\eIVx{
H = {2\pi\over L} \( L_0 + \bar{L}_0 - {c\over 12} \) . }
Using the formula \edet, one obtains:
\eqn\eIVxi{\eqalign{
Z_{h' ,h} &=  \inv{2} \tilde{q}^{-1/48} g_+ (h) g_+ (h')
\prod_{n=0}^\infty \( 1 + a_{n+1/2} (h') a_{n+1/2} (h) ~ \tilde{q}^{n+1/2} \)
\cr
&~~~~~+ \inv{2} {\rm sign}(hh') g_- (h) g_- (h') \tilde{q}^{1/24}
\prod_{n=0}^\infty \( 1 + a_{n} (h') a_n (h) ~ \tilde{q}^{n} \)
,\cr}}
where $\tilde{q} = \exp (-4\pi R/L )$.

The free and fixed boundary conditions at $h=0$ and $h=\pm \infty$
respectively are conformal. In these limits, the above partition function
corresponds precisely to the modular transformation of
the formula \Rchan.
To show this, recall that in
the L-channel, the partition function is\rcardy:
\eqn\Lchan{
Z_{ab} = \sum_{i,j}  n^i_{ab} ~  S^j_i ~ \chi_j (\tilde{q}) ,}
where $S$ determines how the characters transform,
and $\tilde{q} = \exp (-4\pi R/L )$.
As $h\to 0$
($ h\to \pm \infty$), $a_n = 1$ ($a_n = -1$). Also,
\eqn\eIVxii{
\eqalign{
h\to 0 :  ~~~~~~~~~~~&g_+ = \sqrt{2} , ~~~~~g_- = 0  \cr
h\to \pm \infty  :  ~~~~~~~~~~~&g_+ = 1 , ~~~~~g_- = 2^{+1/4} .
\cr}}
The Virasoro characters have the infinite product expressions \gins:
\eqn\eIVxiii{\eqalign{
\chi_0 (q) + \chi_{1/2} (q) &= q^{-1/48} \prod_{n=0}^\infty
\( 1 + q^{n+1/2} \)  \cr
\chi_0 (q) - \chi_{1/2} (q) &= q^{-1/48} \prod_{n=0}^\infty
\( 1 - q^{n+1/2} \)  \cr
 \chi_{1/16}  (q) &=  q^{1/24} \prod_{n=0}^\infty
\( 1 + q^{n} \).  \cr
}}

Let $f$ denote free, and $\pm$ denote the fixed boundary conditions
as $h\to \pm \infty$.  Then in these conformal limits,
from \eIVxi\  one recovers \Lchan, where the only non-zero
$n^i_{ab}$ are $n^0_{\pm\pm} , n^0_{ff} , n^{1/2}_{ff} ,
n^{1/2}_{\pm\mp} $, $n^{1/16}_{\pm f} $, and $S$ is given by
\eqn\ess{
S = \inv{2} \left(\matrix{1&1&\sqrt{2}\cr
1 &1 & -\sqrt{2} \cr
\sqrt{2} & -\sqrt{2} & 0 \cr}\right) . }
(The rows and columns refer to the characters in the order
$0, 1/2, 1/16$.)
For instance,
\eqn\einstance{
Z_{+-} = Z_{h'=\infty , h= -\infty} = \inv{2} \( \chi_0 (\tilde{q} )
+ \chi_{1/2} (\tilde{q} ) - \sqrt{2} \chi_{1/16} (\tilde{q} ) \) . }

\vfill\eject

\centerline{\bf Part II: The sine-Gordon case}

\bigskip

\newsec{TBA in the L channel.}

\subsec{The inhomogeneous 6-vertex model with boundary fields}

Notations closely follow those of \FS. One starts from the inhomogeneous
6-vertex model with boundary fields $h$ as a regularization of the boundary
sine-Gordon model with Dirichlet boundary conditions.

In the inhomogeneous antiferromagnetic 6-vertex model with anisotropy
parameter $\gamma$,
one gives an alternating  imaginary part
$\pm i\Lambda$ to the spectral parameter on alternating vertices
\refs{\DV,\RS}. The scaling limit is given by taking
$\Lambda\rightarrow\infty$, $N\rightarrow\infty$, and the lattice spacing
$\Delta\rightarrow 0$, such that $R\equiv N\Delta$ remains finite. In the bulk,
this provides a regularization of the sine-Gordon model with
Lagrangian
\eqn\sg{L_{\hbox{SG}}
=\int_0^R dx\ \left[ \half (\del\phi)^2 + \mu^2 \cos \beta_{SG}
\phi\right]}
where  $\mu\propto {1\over\Delta}\exp(-{\rm const}\Lambda)$,
$\beta_{SG}^2=8(\pi-\gamma)$, and the field is fixed at $x=0$ and $x=R$
(Dirichlet
boundary conditions) to a value that is simply related to $h$.

The wave function of the inhomogeneous six-vertex model can be expressed in
terms of a set of ``roots''  $\alpha_j$, where $j=1\dots n$. They must be
solutions of the set of equations \refs{\ABBBQ,\skly}
\eqn\foral{\eqalign{N\left[f(\alpha_j+\Lambda,\gamma)+f
(\alpha_j-\Lambda,\gamma)\right]+
2f(\alpha_j,\gamma H)=\cr
2\pi l_j + \sum_{m=1,m\ne
j}^n \left[f(\alpha_j-\alpha_m,2\gamma) +
f(\alpha_j+\alpha_m,2\gamma)\right],\cr}}
where $l_j$ is an integer and all $\alpha_j$ are positive. The function $f$ is
defined as
$$f(a,b)=2\tan^{-1}\left(\cot{b\over 2}\tanh a\right)$$
%
and
\eqn\forH{H\equiv{1\over\gamma} f(i\gamma,-i\ln(h+\cos\gamma)).}
For $h=0$, $\gamma H=\pi-\gamma$. By construction of the Bethe-ansatz wave
function, $\alpha_j>0$. Even though there is a solution of  \foral\ with one
vanishing root for any $N$ and $n$, we emphasize that $\alpha_j=0$ is
{\bf not} allowed  because the wave function vanishes identically in this
case. Observe that equations \foral\ are formally satisfied
as well by the opposite of the roots, $-\alpha_j$. Often in what follows
we shall consider that the roots take both signs in order to rewrite equations
in a way which is similar to the bulk case.

The solutions of these equations are quite intricate for arbitrary $\gamma$
\TS. For simplicity, we restrict to the case $\gamma={\pi\over t}$ where $t$
is an integer, and restrict to the choice $\epsilon=-1$.
In the sine-Gordon model, this falls in the repulsive regime. We make the
standard assumption that all the solutions of interest are collections of
``$k$-strings'' for $k=1,2\dots t-1$ and antistrings $a$
\TS. A $k$-string is a group of $\alpha_j$ in the pattern
$\alpha^{(k)}-i\pi(k-1),  \alpha^{(k)}-i\pi(k-3), \dots, \alpha^{(k)}+
i\pi(k-1)$ where $\alpha^{(k)}$ is real. The antistring has
$\alpha_j=\alpha^{(a)}+i\pi$, where  $\alpha^{(a)}$ is real.

The thermodynamic  limit is obtained by sending $N\to\infty$. In this case,
we can define densities of the different kinds of solutions.
The number of allowed solutions of
\foral\ of type $k$ in the interval $(\alpha,\alpha+d\alpha)$ is
$2 N (\rho_k(\alpha) + \rho_k^h(\alpha)) d\alpha$, where $\rho_k$ is the
density
of ``filled'' solutions (those which appear in the sum in the right-hand-side
of \foral\ ) and $\rho_k^h$ is the density of ``holes'' (unfilled solutions).
The densities $\rho_a$ and $\rho_a^{h}$ are defined likewise for the
antistring. The ``bare'' Bethe ansatz equations follow from taking the
derivative of $\foral$. For $\gamma=\pi/t$ they can be written in the form:
\eqn\bba{\eqalign{2\pi (\rho_k + \rho_k^h)&=  a_k(\alpha) -
\dot{\phi}_{k,t-1}*\rho_{a} +\sum_{l=1}^{t-1}
\dot{\phi}_{kl}*\rho_l +{1\over 2N}u_k\cr
2\pi(\rho_{a} +\rho_a^h)&=2\pi(\rho_{t-1}+ \rho_{t-1}^h) +
{1\over 2N}(u_a-u_{t-1})\cr}}
where $*$ denotes convolution:
$$ f*g(\alpha)\equiv \int_{-\infty}^{\infty} d\alpha' f(\alpha-\alpha')
g(\alpha').$$
These densities are originally defined for $\alpha>0$, but the equations allow
us to define $\rho_k(-\alpha)\equiv\rho_k(\alpha)$ in order to rewrite the
integrals to go from $-\infty$ to $\infty$. The kernels in these equations are
defined most easily in terms of their Fourier transforms
\eqn\fourier{\hat{f}(x)=\int_{-\infty}^\infty {d\alpha\over 2\pi}e^{i\alpha
tx/\pi}
f(\alpha),\quad f(\alpha)={t\over\pi}\int_{-\infty}^\infty e^{-i\alpha
tx/\pi}\hat{f}(x)dx.}
One has
\eqn\forphi{\hat{\dot{\phi}}_{kl}(x)=
\delta_{ab}-2{\cosh x \sinh(t-k)x
\sinh lx \over \sinh  x\sinh tx},}
for $k\geq l$  with $\dot{\phi}_{lk}=\dot{\phi}_{kl}$, and
\eqn\forps{\eqalign{\hat a_k&={\sinh (t-k) x
\over\sinh  tx} \cos\Lambda tx/\pi\cr
\hat u_k&=2 {\sinh (t-H)x \sinh kx  \over \sinh x \sinh tx} +
{\sinh (t-2k)x/2\over \sinh  tx/2} - 1\cr
\hat u_a&=2{\sinh Hx \over \sinh  tx}- {\sinh
(t-2)x/2\over \sinh tx/2} - 1,\cr}}
with in particular $a_1(\alpha)={1\over 2}\left[\dot{f}(\alpha+\Lambda,\gamma)
+\dot{f}(\alpha-\Lambda,\gamma)\right], \phi_{11}(\alpha)=-f(\alpha,2\gamma)$.
The boundary manifests itself in the first term in $u_k$; notice that even for
$h=0$, it still modifies the equations. A few technicalities account for the
other terms (these are relevant here because we are interested in subleading
boundary effects). The second term in $u_k$ arises from the fact that the sum
in \foral\ does not include the term $m=j$; the integration over densities
includes such a contribution and so it must be subtracted off by hand. The
third term in $u_k$ arises because $\rho$ and $\rho^h$ are defined for allowed
solutions, while as already explained, $\alpha=0$ is not allowed because it
does not give a valid wavefunction. Since it is a valid solution of \foral\ but
is not included in the densities, we must subtract an explicit ${2\pi\over
2N}\delta(\alpha)$ (corresponding to $1/2N$ in Fourier space). Explicitly,
one has
\eqn\formi{u_1=2\dot{f}(\alpha,\gamma
H)+2\dot{f}(2\alpha,2\gamma)-2\pi\delta(\alpha).}

For compactness we rewrite \bba\ as
\eqn\newbba{2\pi\epsilon^{(k)}(\rho_k+\rho_k^h)=a_k(\alpha)+\sum_l
\dot{\phi}_{kl}*\rho_l+{1\over 2N}u_k,}
where $\epsilon^{(k)}=-1$ for the antistring.

The energy reads, with proper hamiltonian normalization,
\eqn\ener{{E^{latt}\over 2N}=-{1\over t}\sum_k \int_{-\infty}^\infty
a_k(\alpha)\rho_k(\alpha)d\alpha.}

It is easy to write the thermodynamic Bethe ansatz  for this model. One
finds that the TBA equations, since they are obtained by a variational method,
do not depend on boundary terms, and read as usual
\eqn\TBA{-{2\over t}a_k(\alpha)=T\ln\left(1+e^{\epsilon_k}\right)-
T\sum_l\epsilon^{(l)}{A_{kl}\over 2\pi}*\ln\left(
1+e^{-\epsilon_l}\right),}
where
\eqn\Adef{A_{kl}(\alpha)=2\pi\epsilon^{(k)}\delta_{kl}\delta(\alpha)-
\dot{\phi}_{kl}\, .}
The free energy does depend on the boundary term and reads
\eqn\freeen{\eqalign{F^{latt}=&-TN\sum_k\int_{-\infty}^\infty
\epsilon^{(k)}a_k(\alpha)\ln
\left(1+e^{-\epsilon_k}\right){d\alpha\over 2\pi}\cr
&-
{T\over 2}\sum_k\int_{-\infty}^\infty
\epsilon^{(k)}u_k\ln\left(1+e^{-\epsilon_k}\right){d\alpha\over2\pi}.\cr}}
In the above formulas the temperature $T$ corresponds in the two-dimensional
point of view to having a cylinder of radius $L=1/T$. We can deduce from this
result the ground state energy. Indeed recall that the ground state is
obtained by $\rho_k=0,k\neq 1$ and $\rho^h_1=0$ so
\eqn\gren{E^{latt}=-N\int_{-\infty}^\infty a_1(\alpha)
|\epsilon_1^-|{d\alpha\over 2\pi}-{1\over 2}\int_{-\infty}^\infty
u_1|\epsilon_1^-|{d\alpha\over
2\pi},}
where from \TBA\ we have
\eqn\epsmin{\hat{\epsilon}_1^{\ -}=-{1\over t}{\cos \left({\Lambda
tx/\pi}\right)\over\cosh x}.}
Replacing and  using \forps\ we find
\eqn\energs{\eqalign{E^{latt}_{bulk}=&-{N\over\pi}\int_{-\infty}^\infty
\cos^2\left({\Lambda
tx\over\pi}\right){\sinh (t-1)x\over\cosh x\sinh tx}\cr
E^{latt}_{bdry}=&-{1\over 2\pi}\int_{-\infty}^\infty {\cos(\Lambda tx/
\pi)\over\cosh x}
\left(2{\sinh (t-H)x\over\sinh tx}+{\sinh (t-2)x/2\over\sinh
tx/2}-1\right),\cr}}
where we used the formula
$$
\int_{-\infty}^\infty a(\alpha)b(\alpha)d\alpha=2t\int_{-\infty}^\infty
\hat{a}(x)\hat{b}(-x)dx \, .
$$

\subsec{Boundary energy and entropy of the SG model with Dirichlet boundary
conditions}

In the continuum limit $\Lambda\rightarrow\infty$ the energy contains various
terms. We keep only the finite part which is obtained by closing the
above integrals in the upper half plane and selecting the pole
at $x=i{\pi\over 2}$, leading to
\eqn\contener{\eqalign{E_{bulk}=&R{m^2\over 4}\cot {t\pi\over 2}\cr
E_{bdry}=&-{m\over 2}\left(2{\sin (t-H)\pi/2\over\sin t\pi/2}-\cot
{t\pi\over 4}-1\right)\,,\cr}}
where $m$ is the soliton mass,
\eqn\solmass{m=2e^{-t\Lambda/2}\, .}
All these results trivially generalize to the case of two different
boundary fields by splitting the $H$ dependent terms into the sum
of an $H$ and an $H'$ term. The bulk result agrees with what is obtained
by other methods \ALZunpub,\FSZ. For completeness it might be useful to write
the boundary energy in terms of the parameter $\xi$ of \GZ:
\eqn\fornick{E_{bdry}=-{m\over 2}\left({\cos(t-1)\xi\over\sin t\pi/2}
+{\cos(t-1)\xi'\over\sin t\pi/2}-\cot t\pi/4-1\right).}
As in the bulk case, when $t$ is even, there are additional logarithmic terms.

We can as well take the continuum limit of the lattice Bethe equations to get
the TBA equations for the sine-Gordon model. This is explained in details
in  \FS\ and \FSW. We simply notice here that in the IR limit, all $\epsilon's$
go to infinity so one finds a vanishing
boundary entropy in \freeen.  In the UV limit, it is easy to solve the system
\TBA. One finds
\eqn\sols{\eqalign{1+x_k&={(k+1)^2\over k(k+2)}\cr
1+x_a&=t,\cr}}
where $x_k=e^{-\epsilon_k}$. Now we have the two identities just proved by
inspection
\eqn\identities{\eqalign{\sum_{k=1}^{t-2}\ln{(k+1)^2\over
k(k+2)}&=\ln{2(t-1)\over t} \, ,\cr
\sum_{k=1}^{t-2}k\ln{(k+1)^2\over k(k+2)}&=(t-1)\ln(t-1)-(t-2)\ln t\, .\cr}}
Using the fact that
\eqn\others{\eqalign{\hat{u}_k(0)&=k\left(2-{2\over t}-{2H\over t}\right)
\, ,\cr
\hat{u}_{t-1}(0)&=(t-1)\left(2-{2\over t}-{2H\over t}\right) \, ,\cr
\hat{u}_a(0)&=-2+{2\over t}+{2H\over t}\, ,\cr}}
one finds that
\eqn\res{\sum_k \ln(1+x_k)\int_{-\infty}^\infty u_k(\alpha)=0 \, .}
Hence the boundary entropy is actually the same in the UV and IR limits,
as was argued for a scalar theory \eqqqq.

\newsec{The Destri De Vega equations for the boundary sine-Gordon model}

\subsec{The  DDV equations with boundary conditions}

We now would like to compute the complete Casimir effect in a theory
with boundary. TBA in the R channel is pretty intricate in the non diagonal
case. We use an alternative method elaborated by Destri and De Vega (DDV) in
the periodic case. The lattice regularization plays a crucial role here - it
is actually not known yet how the DDV method is related to the standard
description of the theory using excitations and $S$ matrices. Interestingly,
the result are very close to those of the TBA of scalar theories in the R
channel, and with some more work could probably be considered as a more
rigorous proof of the results presented in the first part.

For that consider eq. \foral\ which we rewrite as
\eqn\newforal{2Np(\alpha_j)+p_{bdry}(\alpha_j)
+\sum_{\alpha_m>0}\phi(\alpha_j-\alpha_m)+\phi(\alpha_j+\alpha_m)=
2\pi n_j,}
where the sum  runs over all roots (including $m=j$) and we introduced the
notations
\eqn\pdef{p(\alpha)\equiv{1\over 2}\left[f(\alpha+\Lambda,\gamma)+
f(\alpha-\Lambda,\gamma)\right],\quad p_H(\alpha)\equiv f(\alpha,\gamma
H),\quad \phi(\alpha)\equiv \phi_{11}(\alpha),}
$$p_{bdry}(\alpha)=2p_H(\alpha)-\phi(2\alpha).$$

The ground state is obtained by filling the real positive solutions,
 $\alpha_j=0$ excepted. This corresponds to the choice $n_j=1,2,\ldots$.
Recall that
if $\alpha_j$ is solution of
\newforal\ with some $n_j$, so is formally $-\alpha_j$ (with $-n_j$).
Given the set of roots $\{\alpha_j>0\}$ representing the ground state, one can
construct the counting function as follows:
\eqn\deff{f(\alpha)\equiv 2iNp(\alpha)+
ip_{bdry}(\alpha)
+i\sum_{\alpha_m>0}\phi(\alpha-\alpha_m)+\phi(\alpha+\alpha_m),}
Define then
\eqn\defy{Y(\alpha)\equiv e^{f(\alpha)}.}
We have $Y(\alpha_j)=Y(-\alpha_j)=1$ for every root $\alpha_j$ from
the ground state, as well as $Y(0)=1$.
 Therefore we
can rewrite \deff\ as
\eqn\newwforal{f(\alpha)=2iNp(\alpha)+
ip_{bdry}(\alpha)-i\phi(\alpha)-
\int_C\phi(\alpha-\alpha'){\dot{Y}(\alpha')\over
1-Y(\alpha')}{d\alpha'\over 2\pi},}
where $i\phi(\alpha)$ at the right-hand side takes care of the unwanted
contribution of the pole $\alpha'=0$
%
%
and the contour $C$ consists of two parts as shown in figure 3, $C_1$ above
 and $C_2$ below the real axis.

\midinsert
\epsfxsize = 3in
\bigskip\bigskip\bigskip\bigskip
\vbox{\vskip -.1in\hbox{\centerline{\epsffile{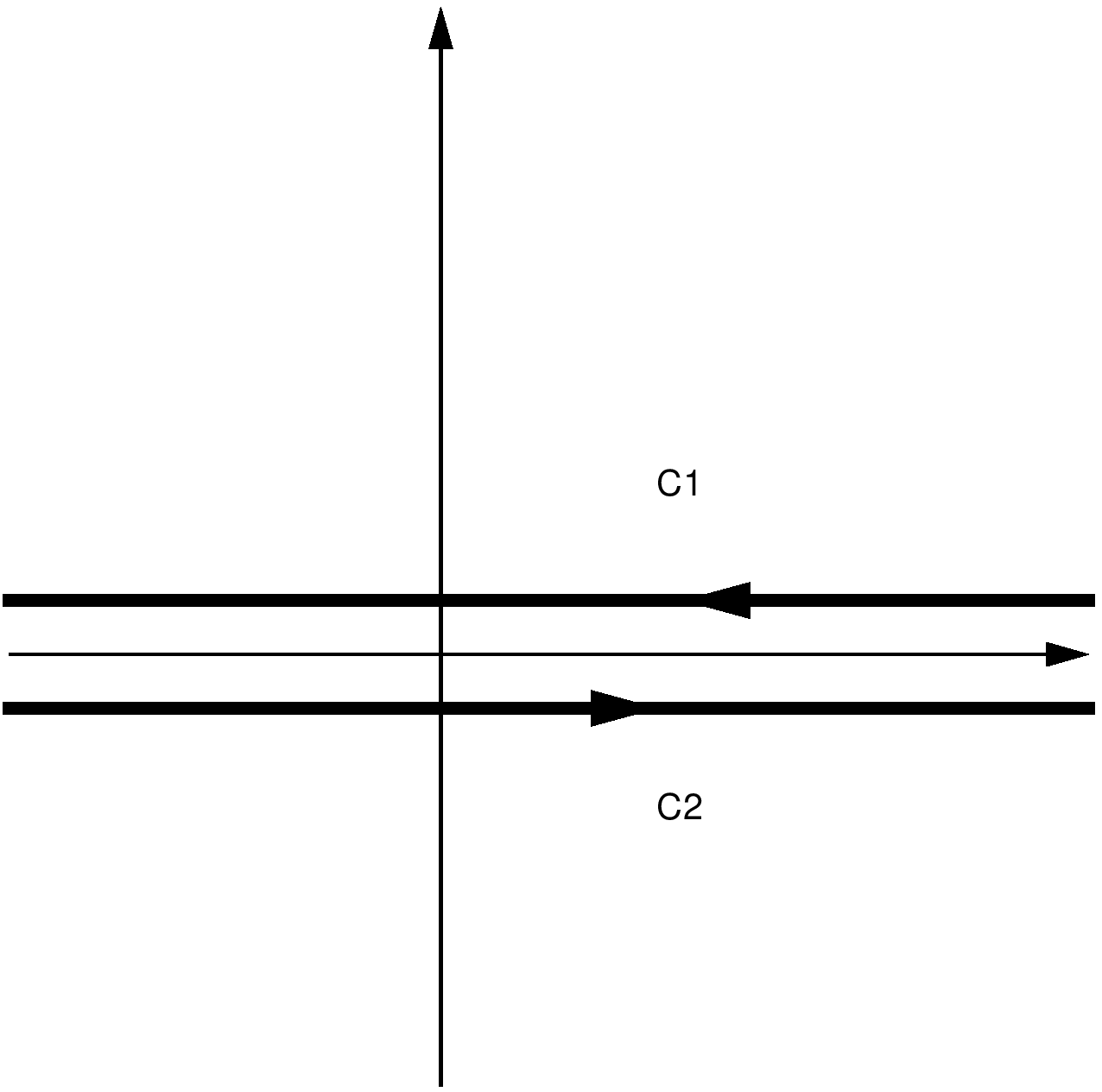}}}
\vskip .1in
{\leftskip .5in \rightskip .5in \noindent \ninerm \baselineskip=10pt
Figure 3.
Contours of integration in the DDV method.
\smallskip}} \bigskip
\endinsert

 Like in the bulk case \DDV,
 simple manipulations
allow us to rewrite this non-linear integral equation as
\eqn\ddvi{\eqalign{f(\alpha)&=2iNP(\alpha)+iP_{bdry}(\alpha)+
\int_{C_1}\Phi(\alpha-\alpha')\ln\left(1-e^{f(\alpha')}\right)
d\alpha'\cr
&+\int_{C_2}\Phi(\alpha
-\alpha')\ln\left(1-e^{-f(\alpha')}\right)d\alpha'.\cr}}
In  equation \ddvi\ one has
\eqn\phiren{\Phi(\alpha)=-
{t\over 2\pi^2}\int_{-\infty}^{+\infty}
dx e^{-itx\alpha/\pi}{\sinh(t-2)x\over 2\sinh(t-1)x\cosh x},}
together with
\eqn\pren{P(\alpha)=
\int_{-\infty}^{+\infty}dx{e^{-itx\alpha/\pi}-1\over -ix}
{\cos(\Lambda tx/\pi)\over 2\cosh x}}
and
\eqn\pbdrren{\eqalign{
P_{bdry}&=
\int_{-\infty}^{+\infty}dx{e^{-itx\alpha/\pi}-1\over -ix}
\left[{\sinh (t-H)x\over\sinh (t-1)x\cosh x}\right.\cr
&\left.+{\sinh (t-2)x/2\cosh tx/2\over\sinh (t-1)x\cosh x}+
{\sinh (t-2)x\over 2\sinh (t-1)x\cosh x}\right],\cr}}
Obtaining \ddvi\ requires some care with the definition of logarithms.
One proceeds as follows. Before integrating by parts in the integral over
$C_2$ one factors out ${d\over dx}\ln Y(x)$:
$$
-{\dot{Y}(x)\over 1-Y(x)}={d\over dx}\ln[1-Y^{-1}] +
{d\over dx}\ln Y(x).
$$
Then both integrals over $C_1$ and $C_2$ can be taken by parts, resulting in
$$
\eqalign{f(\alpha)&-\int_{-\infty}^{+\infty}\dot{\phi}(\alpha-\alpha')
f(\alpha'){d\alpha'\over 2\pi}=2iNp(\alpha)+
ip_{bdry}(\alpha)-i\phi(\alpha)\cr
&-\int_{-\infty}^{+\infty}\dot{\phi}(\alpha-\alpha'-i0)\ln[1-Y(\alpha'+i0)]
{d\alpha'\over 2\pi} \cr
&+\int_{-\infty}^{+\infty}\dot{\phi}(\alpha-\alpha'+i0)
\ln[1-Y^{-1}(\alpha'-i0)]
{d\alpha'\over 2\pi}
,\cr}
$$
(surface terms from $C_1$ and $C_2$ cancel against each other provided
$N,\Lambda$ are finite).
To make the source term vanish at infinity  we
take the derivative of both sides in the latter equation, after which it
can be Fourier-transformed and ``dressed'' by the factor
$(1-\hat{\dot{\phi}})^{-1}$. Finally, one goes back in the rapidity space
and integrates the equation, using $f(0)=0$, to obtain \ddvi.

The energy of the ground state configuration can be expressed as
\eqn\enerix{\eqalign{E^{latt}&=-{2\over t}\sum_{\alpha_j>0} \dot{p}(\alpha_j)=
-{1\over t}\sum_{\alpha_j}\dot{f}(\alpha_j-\Lambda, \gamma)+
\dot{f}(-\alpha_j-\Lambda, \gamma)\cr
&=
{1\over t}\dot{f}_{\gamma}(\Lambda)+
{1\over t}\int_C \dot{f}_{\gamma}(\alpha-\Lambda)
{\dot{Y}(\alpha)\over 1- Y(\alpha)}
{d\alpha\over 2i\pi},\cr}}
where $f(\alpha, \gamma)\equiv f_{\gamma}(\alpha)$.
One finds after exactly the same manipulations as above,
\eqn\tati{
\eqalign{&E={1\over t}\dot{f}_{\gamma}(\Lambda)-
{1\over t}\int_{-\infty}^{+\infty}f_{\gamma}''(\alpha-\Lambda+i0)
\ln[1-Y(\alpha+i0)]{d\alpha\over 2i\pi}\cr
&+{1\over t}\int_{-\infty}^{+\infty}f_{\gamma}''(\alpha-\Lambda-i0)
\ln[1-Y^{-1}(\alpha-i0)]{d\alpha\over 2i\pi}
+{1\over t}\int_{-\infty}^{+\infty}f_{\gamma}''(\alpha-\Lambda)f(\alpha)
{d\alpha\over 2i\pi}.\cr}}
Substituting \ddvi\ instead of $f(\alpha)$ in the last term of the latter,
we obtain
\eqn\eneriixx{\eqalign{E^{latt}=&E^{latt}_{bulk}+E^{latt}_{bdry}
 - {i\over t}\int_{-\infty}^{+\infty}
s(y-\Lambda+i0)\ln[1-Y(y+i0)]{dy\over 2\pi}\cr
&+{i\over t}\int_{-\infty}^{+\infty}
s(y-\Lambda-i0)\ln[1-Y^{-1}(y-i0)]{dy\over 2\pi},\cr}}
where we defined $s(y)$ by
\eqn\epsdefx{s(y)=i\int_{-\infty}^{+\infty}{kdk\over 2\cosh\gamma k}
e^{-iky}={t^2\over 4}{\tanh(ty/2)\over\cosh(ty/2)}.}
The last two terms in \eneriixx\ represent finite-size corrections to
the ground state energy, while
\eqn\kust{\eqalign{E^{latt}_{bulk}+E^{latt}_{bdry}=&{1\over
t}\dot{f}_{\gamma}(\Lambda)
-{1\over t}\int_{-\infty}^{+\infty}{dk\over 2\pi}
e^{-ik\Lambda}\hat{\dot{f}}_{\gamma}(k){2N\hat{\dot{p}}(k)+\hat{\dot{p}}_{bdry}
(k)-
\hat{\dot{\phi}}(k)\over 2\pi-\hat{\dot{\phi}}(k)}\cr \equiv
& {1\over t}\dot{f}_{\gamma}(\Lambda) - {2N\over t}\int_{-\infty}^{+\infty}
\dot{f}_{\gamma}(\alpha-\Lambda)\rho(\alpha)d\alpha, \cr}}
where the function $\rho(\alpha)$ defined so satisfies the following equation:
\eqn\kustik{2\pi\rho(\alpha)=\dot{p}(\alpha)+\dot{\phi}\ast\rho(\alpha)
+{1\over 2N}\left(\dot{p}_{bdry}(\alpha)-\dot{\phi}(\alpha)\right),
}
which can be checked by solving this linear equation in Fourier space.
Introduce $\rho_1(\alpha)=\rho(\alpha)-\delta(\alpha)/2N$.
Then
\eqn\dela{ E^{latt}_{bulk}+E^{latt}_{bdry}=- {2N\over
t}\int_{-\infty}^{+\infty}
\dot{f}_{\gamma}(\alpha-\Lambda)\rho_1(\alpha)d\alpha}
and, by virtue of \kustik,  $\rho_1$ satisfies the equation
\eqn\kustiki{2\pi\rho_1(\alpha)=\dot{p}(\alpha)+\dot{\phi}\ast\rho_1
(\alpha)
+{1\over 2N}\left(\dot{p}_{bdry}(\alpha)-2\pi\delta(\alpha)\right).}
Hence $\rho_1$ is the density of the ground state configuration \bba\
(see also \formi) and $E^{latt}_{bulk}$
and $E^{latt}_{bdry}$
coincide with the quantities computed in the previous section.

\subsec{The continuum DDV equations}

Having checked the correct values of bulk and boundary energies,
we now let the cutoff $\Lambda\to\infty$ according to $\Lambda = {2\over t}
\log (2/m\Delta)$ with  the size of the system $R=N\Delta$ and the   physical
mass $m={2\over\Delta}e^{-t\Lambda/2}$ fixed, and work only with
the renormalized theory. In that limit one has, evaluating all integrals
in Fourier transforms and keeping the leading terms,
\eqn\conteqs{s(\alpha+\Lambda)+s(\alpha-\Lambda)=
-{t^2\over 2}m\sinh\theta,\qquad
NP(\theta)=mR\sinh\theta,}
where we set
\eqn\defrap{\theta={t\alpha\over 2}.}
Recall indeed that when one studies the excitations of the model,
a relativistic dispersion relation is obtained provided
the rapidity in the relativistic theory and the ``bare rapidity'' of the
Bethe excitations are related by \defrap. We redefine implicitely all functions
to depend on $\theta$ from now on. The energy therefore reads now,
\eqn\newenrform{\eqalign{E=&E_{bulk}+E_{bdry}+
{1\over 2}\int_{C_1} m\sinh\theta \ln\left(1-e^{
f(\theta)}\right){d\theta\over
2i\pi}\cr
&+{1\over 2}\int_{C_2} m\sinh\theta \ln\left(1-e^{
-f(\theta)}\right){d\theta\over
2i\pi},\cr}}
where $f$ is solution of the integral equation
\eqn\renormbethe{\eqalign{f(\theta)=&2imR\sinh\theta +i
P_{bdry}(\theta)+\int_{C_1}\Phi(\theta-\theta')
\ln\left(1-e^{f(\theta')}\right){d\theta'}\cr
+&\int_{C_2}\Phi(\theta-\theta')
\ln\left(1-e^{-f(\theta')}\right){d\theta'},\cr}}
and
\eqn\kerndef{\Phi(\theta)=-\int_{-\infty}^\infty
{dx\over 2\pi^2}{\sinh (t-2)x\over
\cosh x\sinh(t-1)x}e^{2ix\theta/\pi},}
which can be identified with
\eqn\ident{\Phi(\theta)=-{1\over 2i\pi}{d\over d\theta}\ln S_{++}(\theta),}
where $S_{++}$ is the soliton-soliton $S$ matrix element.
We have also from \pbdrren\ and the identity $(18)$ in \FS\ (see also \GNM)
\eqn\pexpl{P_{bdry}(\theta)=-{1\over i}\ln S_{++}(2\theta) +{2\over i}
\ln {\cal R}_{++}(\theta)+\pi,}
(so ${d\over d\theta}P_{bdry}=\Theta+2\pi\delta(\t)$). The additional
$\pi$ in the above equation arises because the $S_{++}$ has
an overall minus sign \ZZ\
(eg $S=-1$ in the free fermion case  $t=2$).

\subsec{The Casimir effect}

We define the effective  central charge by the formula
\eqn\cdeff{E=E_{bulk}+E_{bdry}-{\pi c_{eff}\over 24R}.}
Observe that in the approach of the first part,
we dealt with a renormalized theory with the vacuum energy subtracted. The
last term in \cdeff\ is a renormalized energy, and it is what we will
compare later to the results of the first part. From \newenrform\ we find
\eqn\cform{c_{eff}(mR)=-{6 R\over i\pi^2}\left\{\int_{C_1}m\sinh\theta\ln\left(
1-e^{f(\theta)}\right)d\theta+\int_{C_2}m\sinh\theta\ln\left(
1-e^{-f(\theta)}\right)\right\}d\theta.}
To study the ultraviolet behaviour of the above expression, let us
use $f(\overline{\theta})=\overline{f(\theta)}$ and rewrite
 \renormbethe\ as
\eqn\rewrit{f(\theta)=2imR\sinh\theta+iP_{bdry}(\theta)-
2i\int_{-\infty}^\infty \Phi(\theta-\theta')\hbox{ Im}
\ln\left[1-e^{f(\theta'+i0)}\right]d\theta',}
and \cform\ as
\eqn\rewriti{c_{eff}(mR)={12mR\over\pi^2}\int_{-\infty}^\infty
d\theta\sinh\theta\hbox{ Im}
\ln\left[1-e^{f(\theta+i0)}\right].}
It might be useful to remind the reader of the form of the corresponding
bulk equations \DDV, \AlZ:
\eqn\rewritibulk{c_{eff}(mR)={6mR\over\pi^2}\int_{-\infty}^\infty
\sinh\theta\hbox{ Im}
\ln\left[1+e^{f(\theta+i0)}\right]d\theta,}
with $f$ satisfying
\eqn\rewritbulk{f(\theta)=imR\sinh\theta+i\omega-2i\int_{-\infty}^\infty
\Phi(\theta-\theta')\hbox{ Im}
\ln\left[1+e^{f(\theta'+i0)}\right]d\theta',}
and $\omega$ is the twist of the 6-vertex model. Note the deep similarity
between these two systems; the factors of $2$ can obviously be
absorbed in a redefinition of $mR$ and the minus sign
in the arguments of logarithms in a redefinition of $f$, so the only
essential difference
is that the twist angle $\omega$ is replaced by $P_{bdry}$. It is well known
that the twist corresponds to a soliton fugacity \FS\ so we recover the result
of the first part  that the boundary acts by some effective, rapidity dependent
fugacity.

In the limit when $R\rightarrow 0$, only the region $|\theta|$ large
contributes to $c_{eff}$. Let us focus on the limit  $\theta>>1$, the results
for negative
$\theta$ following by symmetry. Then one finds $f\approx f_K$, where
\eqn\bethekink{f_K(\theta) =imR e^\theta+iP_{bdry}(\infty)
-2i\int_{-\infty}^\infty \Phi(\theta-\theta')\hbox{ Im}
\ln\left[1-e^{f_K(\theta'+i0)}\right]d\theta',}
together with
\eqn\uvc{c_K(mR)= {6\over\pi^2}\int_{-\infty}^\infty mRe^{\theta}
\hbox{ Im} \ln\left[1-e^{f_K(\theta+i0)}\right]d\theta.}
It is now useful to recall some well known results about
dilogarithms \KR. Define
\eqn\dilogdef{L(x)\equiv \int_0^x du\left[{\ln(1+u)\over u}-
{\ln u\over 1+u}\right].}
Assume
\eqn\suppi{-i\ln F(x)=\phi(x)+2\int_{-\infty}^\infty dy \ G(x-y)\hbox{ Im}
\ln\left[
1+F(y+i0)\right],}
with  $G$ an even function. Then one has
\eqn\main{\eqalign{\hbox{Im}\int_{-\infty}
^\infty dx\phi'(x)\ln\left[1+F(x+i\epsilon)\right]={1\over 2}\hbox{Re}\left\{
L[F(-\infty)]-L[F(\infty)]\right\}\cr
+{1\over 2}\hbox{Im}\left\{
\phi(\infty)\ln\left[1+F(\infty)\right]-\phi(-\infty)
\ln\left[1+F(-\infty)\right]
\right\},\cr}}
(where we did not write the $i0$ part of some arguments for
simplicity). Set $F=e^{f_K-i\pi}$ and denote
$P_{bdry}(\infty)\equiv\sigma$. Then, according to \bethekink:
 $\phi=mRe^\theta+\sigma-\pi, \quad G=-\Phi$.
We have $\phi(-\infty)=\sigma-\pi$, $\phi(\infty)=\infty$.
One has also $F(\infty+i0)=0$, and from this and \bethekink\ one can get
the value of $F$ at $-\infty$:
\eqn\fff{\eqalign{{f_K(-\infty)\over i}=&
\sigma-2\hbox{ Im}\ln\left[1-e^{f_K(-\infty+i0)}\right]
\int_{-\infty}^\infty \Phi(\theta)d\theta \cr
=&\sigma+{t-2\over t-1}\hbox{ Im}\ln\left[1-e^{f_K(-\infty+i0)}\right].\cr}}
So, if $e^{f_K(-\infty+i0)}=e^{i\omega}$ one may use
$\hbox{ Im}\ln(1\pm e^{i\omega})= {1\over 2i}\ln(\pm e^{i\omega})$ to   find
\eqn\omm{e^{i\omega}=
-\exp\left\{2i{t-1\over t}\sigma+2i{\pi\over t}\right\}.}
 From \main\ it follows that
$$\hbox{ Im}\int_{-\infty}^\infty
mRe^{\theta}\ln(1-e^{f_K(\theta+i0)})d\theta={1\over 2}
\hbox{ Re}L(-e^{i\omega})-{1\over 2}\left(\sigma-\pi\right)
\left({t-1\over t}\sigma-\pi+{\pi\over t}\right).$$
In the region $\theta<<1$ we have $f_1\approx f_A$, and similar calculations
yield
$$\hbox{ Im}\int_{-\infty}^\infty
mRe^{-\theta}\ln(1-e^{f_A(\theta+i0)})d\theta=-{1\over 2}
\hbox{ Re}L(-e^{-i\omega})+{1\over 2}\left(\pi-\sigma\right)
\left(-{t-1\over t}\sigma+\pi-{\pi\over t}\right).$$
Collecting both $\theta>>1$ and  $\theta<<1$ contributions we obtain
\eqn\ccc{\eqalign{c_{UV}=&{6\over\pi^2}\left\{{1\over 2}
\left[L(-e^{2i\omega)})+
L(-e^{-2i\omega})\right]+
 {t-1\over t}\left(\sigma-\pi\right)^2\right\}=\cr
=&{6\over\pi^2}
\left[{\pi^2\over 6}-{t-1\over
t}\left(\sigma-\pi\right)^2\right].\cr}}
{}From \pbdrren\ we get
\eqn\pinf{P_{bdry}(\infty)\equiv\sigma= 2\pi-{\pi\over 2}{H+H'\over t-1}.}
Finally, from this we find
\eqn\cccc{c_{UV}=1-6{t-1\over t}\left(1-{H+H'\over 2(t-1)}\right)^2.}
In the  case with no boundary field, $H=H'=t-1$ so $c_{UV}=1$ as expected.

\vskip1cm

{\it Remark.} So far we tacitly assumed that $0<H<t-1$. In general,
from relation \forH\ follows that when $h>0, \quad H$ varies between $-t-1$
and $-1$, while when $h<0, \quad -1<H<t-1$. To generalize our results,
we should use the most general form of $p_H$:
%
\eqn\balda{\hat{\dot{p}}_H(k)=\int_{-\infty}^\infty\dot{p}_H(\alpha)e^{ik\alpha}d\alpha=
2\pi
\hbox{ sign}(H){\sinh(\pi-\omega_H)k\over\sinh\pi k},\quad -\pi<\gamma H<\pi,}
where we defined $\omega_H\equiv|\gamma H|$. For $-2\pi<\gamma H<-\pi$
set $\omega_H=2\pi+\gamma H$ and $\hbox{ sign}(H)=1$ in \balda.
Then \pinf\ generalizes to
$$
\sigma=2\pi-{\pi\over 2}{\omega_H+\omega_H'\over \pi-\gamma}
$$
if $H$ and $H'$ are both positive or $-2t<(H, H')<-t$, and
$$
\sigma=2\pi-{\pi\over 2}{4\pi-\omega_H-\omega_H'\over \pi-\gamma}
$$
if they are both negative, but greater than $-t$.
In the case when $0<H<t-1$ and $-t<H'<-1$  (that is, $h<0, h'>0$) we get:
$$
\sigma=\pi+{\pi\over 2}{\omega_H'-\omega_H-2\gamma\over \pi-\gamma}.
$$
So, when $\omega_H'-\omega_H=2\gamma$ we have $\sigma=\pi$ and $c_{UV}=1$,
as in the free case. The condition $\omega_H'-\omega_H=2\gamma$
is equivalent to $h=-h'$, as could be seen from \forH. That $c=1$ when the
two surface field are real and opposite is well known from lattice studies \BS.

\newsec{Remarks on the DDV equations for minimal models and excited states}

A particularly interesting case is when the XXZ chain or the inhomogeneous
6-vertex model commutes with the quantum group $U_qsl(2)$. In that
case $h=-h'=2i\sin\gamma$ and the net result is that all $H$ dependent terms
simply disappear from the equations so in particular
\eqn\equu{E_{bdry}^{qu}={m\over 2}\left(\cot{t\pi\over 4} +1\right).}
At the $N=2$ supersymmetric point, $t=3$ so the boundary energy vanishes,
a result
well expected from supersymmetric considerations. More generally, it vanishes
if $t=4n+3$, $n$ an integer. Notice that the bulk energy vanishes
for $t$ an odd number (as
a consequence of the generalized fractional $N=2$ supersymmetries
studied in \ref\vafa{A. LeClair and C. Vafa, Nucl. Phys. B401 (1993) 413. }).

In the quantum group symmetric case \PS\ one has $H+H'=2t$ so from \cccc\
\eqn\cmin{c=1-{6\over t(t-1)},}
the expected result for the restricted sine-Gordon model\ressmir\ref\restr{A.
LeClair, Phys. Lett. 230B (1989) 103.}\ref\resden{D. Bernard and
A. LeClair, Nucl. Phys. B340 (1990) 409.}.

In the quantum group case, all $H$-dependent terms simply disappear
from the equations (in notations of \GZ\ this corresponds to
$\xi\to\pm i\infty$). We find then the value of the boundary matrix element
\eqn\quantumbdry{\ln {\cal R} = {1\over i}\int_{-\infty}^\infty
{dx\over x}{\sinh (3x/2)\sinh (t-2)x/2
\over \sinh (t-1)x/2\sinh (t-1)x}\sin {2\over \pi} x\t\, .}
In the last equation we suppressed the label $++$ because the scattering of
solitons and antisolitons can be made identical by the same change of gauge
which ensures commutation of the bulk S-matrix with $U_qsl(2)$. To understand
\quantumbdry, we can use the reformulation of the quantum group
symmetric bulk-scattering in terms of the $O(n)$ model.
In that case, the bulk S-matrix can be rewritten \Sasha\
\eqn\bulkS{S_{ab}^{cd}={\rho(-i\t)\over i}\left[
\sinh\lambda (i\pi-\t) \,\delta_a^b\delta_c^d + \sinh\lambda\t \,
\delta_{ac}\delta^{bd}\right],}
where
\eqn\pref{\rho(-i\t) =
-i \sinh^{-1}\lambda (i\pi - \t) \,
\exp\left[{1\over i}\int_0^{\infty} {dx \over x}
{\sinh\left(  \left({1\over\lambda}-1\right)x\right)
 \over \sinh\left({x\over\lambda}\right)
\cosh\left(x\right)}\sin {2\over\pi}x\t\right] . }
The labels correspond to colors in an $O(n)$ symmetric model,
running formally  $a=1,\ldots,n$. One has  $n=2\cosh i\pi\lambda$
so  the sine-Gordon model
corresponds to $n<2$ and one must continue things
analytically in $n$. In our previous notations, $\lambda={1\over t-1}$. In this
approach, the particles are self conjugate so the unitarity
crossing relation reads
\eqn\nunitcross{{\cal R}_a^a\left(i{\pi\over 2}-\t\right) = S_{aa}^{bb}(2\t)
{\cal R}_b^b\left(i{\pi\over 2}+\t\right).}
We deal with the $O(n)$ model with free boundary conditions so by $O(n)$
symmetry, ${\cal R}_a^a$ actually does not depend on $a$ let us just call it
${\cal R}$. Then \nunitcross\
reads
\eqn\newversi{\eqalign{{\cal R}\left(i{\pi\over 2}-\t\right)& =
{\rho(-2i\t)\over i}
\left[ n\sinh 2\lambda\t+\sinh\lambda(i\pi-2\t)\right]
{\cal R}\left(i{\pi\over 2}+\t\right)\cr
&=\sin\lambda(\pi-2i\t)\rho(-2i\t)
{\cal R}\left(i{\pi\over 2}+\t\right),\cr}}
where we used $n=2\cosh i\pi\lambda$. The latter equation
appear in \GZ, and its minimal solution is precisely \quantumbdry.
More results about the boundary $O(n)$ model will
be presented elsewhere, together with a kink interpretation
for minimal models.

Now in \BS,
it was shown how one can obtain the various conformal weights $h_{rs}$ by
changing the boundary conditions and the total value of the spin in the XXZ
chain. The change of boundary conditions proposed in \BS\ was to project the
first $r$ spins near a boundary onto the spin $r$ representation of
$U_qsl(2)$, which leads to characters $h_{r,1}$. Now let us recall
(see \SS\ for discussion, see also \NM) that if ${\cal R}_{++},{\cal R}_{--}$
are boundary Reflection-matrices satisfying unitarity and crossing, one can
obtain new solutions of the same equations by defining
\eqn\defdef{\eqalign{{\cal R}^{(2)}_{++}(\t)&\equiv a(\t-\t_0) a(\t+\t_0)
{\cal R}_{++}(\t)\cr
{\cal R}^{(2)}_{--}(\t)&\equiv b(\t-\t_0) b(\t+\t_0) {\cal R}_{--}(\t) +
c(\t-\t_0) c(\t+\t_0) {\cal R}_{++}(\t),\cr}}
where $a,b,c$ denote the standard elements of the bulk sine-Gordon S-matrix,
and $\t_0$ is a free parameter. The process can be iterated again to give
R-matrices
${\cal R}^{(n)}$. Now one can again let $\xi\to\pm i\infty$ to
obtain what we conjecture to be the continuum limit of the projection discussed
in \BS. The DDV equations read as above, but with ${\cal R}^{(r)}$ replacing
${\cal R}$. The effective central charge follows from $P_{bdry}(\infty)$. Using
$$
{2\over i}
\ln S_{++}(\infty)=-\pi {t\over t-1}
$$
one finds
\eqn\newpbbb{P_{bdry}(\infty)=\pi\left({t-2\over t-1}-(r-1){t\over
t-1}\right).}
Using the equations \fff\ to \cmin\ this leads to
\eqn\ceee{c_{eff}^{(r)}=1-6{t-1\over t}\left(1-{t\over
t-1}r\right)^2=c-24h_{r1}.}
Hence we believe we can observe the conformal weights $h_{r1}$ using these
``boundary fused'' R-matrices (what value to give to $\t_0$ requires more
discussion).

Changing the value of $s$ is more difficult from the DDV point of view.
On the lattice, all one has to do is to look at non vanishing spin $m$. Then
one observes $h_{r,1+2m}$, corresponding to
\eqn\hopedfor{P_{bdry}(\infty)=\pi\left(-{1\over t-1}-(r-1){t\over
t-1}+(s-1)\pi
\right).}
Unfortunately all our equations are defined modulo $2\pi$. So it is possible to
observe $h_{r,2}$ only. For that purpose simply subtract $\pi$ to \pexpl (this
corresponds to having a lattice with an odd number of sites).

\newsec{The massless DDV equations}

The structure of the DDV equations is pretty transparent in continuum terms:
the function $\Phi$ takes care of the interaction between solitons and
antisolitons, and the term $P_{bdry}$ takes care of all the boundary effects.
It is then natural to guess what the DDV equations would look like in other
cases. A case of special interest is the bulk massless case. If the boundary
is also conformal invariant, one gets the set of equations \bethekink\ and
\uvc, and we see that the particular conformal boundary condition is fully
characterized by {\bf a rapidity independent phase}. If the boundary is
perturbed and the perturbation is integrable, its action can be described by
a boundary scattering. If moreover this scattering is diagonal and identical
for solitons and antisolitons, it is easy to write the equations as
\eqn\massless{\eqalign{f(\theta) =&i mR e^\theta+iP_{bdry}(\t-\t_B)
-2i\int_{-\infty}^\infty \Phi(\theta-\theta')\hbox{ Im}
\ln\left[1-e^{f(\theta'+i0)}\right]d\theta'\cr
c(me^{\t_B}R)=& {12\over\pi^2}\int_{-\infty}^\infty mRe^{\theta}
\hbox{ Im} \ln\left[1-e^{f(\theta+i0)}\right]d\theta.\cr}}
Here $\t_B$ characterizes the strength of the boundary perturbation, and
$me^{\t_B}$ is the usual ``Kondo'' or ``boundary'' temperature. The only
$\t$ dependent term in $P_{bdry}$ is the one involving the boundary S-matrix.
The simplest case corresponds to the following boundary S-matrix
\eqn\sexem{{\cal R}(\t-\t_B)=i\tanh
\left({\t-\t_B\over 2}-{i\pi\over 4}\right)\, ,}
corresponding to an interaction of anisotropic Kondo type on one side
of the strip, and we chose the usual Dirichlet boundary conditions
on the other side. Then one has
\eqn\pbdrmassless{P_{bdry}(\t-\t_B) = {1\over i}\ln
\tanh \left({\t-\t_B\over 2}-{i\pi\over 4}\right).}
As a result, the energy scales as the ground state of the free boson
theory with $c=1$ in the UV, but according to \cccc\ in the IR one observes an
excited state of conformal weight
\eqn\weight{h={t-1\over 4t}\left({P_{bdry}(\infty)\over\pi}\right)^2={t-1\over
4t}.}
Microscopically, the IR limit corresponds to having an infinite boundary field
in the XXZ chain and one easily checks directly the validity of \weight.

\newsec{Conclusions}

To conclude, consider the energy of the renormalized theory,
$E=-{\pi c_{eff}\over  24R}$. So far,
 the relation between the DDV equations and the underlying
scattering structure is unknown. Nevertheless, it is likely that the two terms
($C_1$ and $C_2$) in the above equations correspond to the existence
of ``soliton'' and ``antisoliton''. By naive extension, the result
for a scalar theory should have the same form but with only one of
these terms. After shifting the countour $C_1$ to
$\hbox{Im}(\theta)=-i\pi/2$ and changing a few notations, one is thus
led to the following conjecture for the ground state energy of a
{\bf scalar} theory with bulk scattering given by $\Phi$:
\eqn\conji{E=-{m\over 4\pi}\int_{-\infty}^\infty
\cosh\theta \ln\left[1+e^{-\epsilon(\theta)}\right]d\theta,}
where now
\eqn\conjii{\epsilon(\theta)=2mR\cosh\theta-iP_{bdry}(\theta+i\pi/2)+i\pi
+\int_{-\infty}^\infty \Phi(\theta-\theta')\hbox{Im}\ln\left[
1+e^{-\epsilon(\theta')}\right]d\theta',}
and
\eqn\phhhh{\Phi(\theta)=-{1\over 2i \pi}{d\over d\theta}\ln S(\theta).}
Observe that, introducing the notation
 $K(\theta) = {\cal R}\left({i\pi\over 2}-\theta\right)$ we can rewrite
\eqn\rewriteit{i\pi-iP_{bdry}(\theta+i\pi/2) =
-\ln K(-\theta)K'(-\theta) -\ln S(2\theta),}
where we used the relations $S(\theta)S(-\theta)=1$ and $S(i\pi-\theta)=
S(\theta)$. We can also use the boundary crossing relation \GZ\
\eqn\bdrycross{K(\theta) = S(2\theta)K(-\theta),}
together with $\bar{K}(\theta) = K(-\theta)$ to rewrite
\eqn\newpbdr{-iP_{bdry}(\theta+i\pi/2)+i\pi=-\ln \bar{K}(\theta)-\ln
K'(\theta).}
This is exactly the equation we deduced in section 3.

\bigskip\bigskip

\noindent{\bf Acknowledgments}: This work was supported
by the Packard Foundation, the National Young Investigator Program
(NSF-PHY-9357207 and NSF-PHY-9357480)
and DOE (DE-FG03-84ER40168). G.M. thanks the
Department of Physics of the University of Southern California and
Newman Laboratory of Cornell University for their hospitality.
A.L. thanks S. Lukyanov for helpful discussions.  H.S. thanks
Al. Zamolodchikov for his kind hospitality in Montpellier where this work
was started, and for an illuminating explanation of the work of Destri and
de Vega in the periodic case. Our formulas for the boundary energy of a
scalar theory were originally presented (without proof) by A.B. Zamolodchikov
at the conference SMQFTS in May 1994. Our work was started earlier, and
although we derived the formulas in a totally independent fashion, we certainly
benefited from this presentation. In this respect, our main contribution
is rather in the second part, where the use of the DDV method allows a more
rigorous derivation, avoiding in particular the uncomfortable use of
``$\delta(0)$''.

\listrefs
\bye